\documentstyle[12pt,eqsection,cite,subeqnarray]{article}

\begin{document}

\title{\vspace*{-2.5cm}Correction-to-Scaling Exponents\break
                     for Two-Dimensional Self-Avoiding Walks}

\author{
  \\
  \hspace*{-6mm}
  {\rm Sergio Caracciolo$^{\rm a}$, 
      Anthony J. Guttmann$^{\rm b}$, Iwan Jensen$^{\rm b}$,} \\
  {\rm Andrea Pelissetto$^{\rm c}$, Andrew N.  Rogers$^{\rm b}$,
       and Alan D. Sokal$^{\rm d}$}
    \\[2mm]
  \hspace*{-4mm}
  {\small\it ${}^{\rm a}$
      Dip. di Fisica and INFN, Universit\`a di Milano, via Celoria 16,
      I-20133 Milano, ITALY}
        \\[-1mm]
  \hspace*{-4mm}
  {\small\it ${}^{\rm b}$
        Dept.~of Mathematics and Statistics, University of Melbourne,
      Vic. 3010, AUSTRALIA}
        \\[-1mm]
  \hspace*{-4mm}
  {\small\it ${}^{\rm c}$
      Dip.~di Fisica and INFN--Sezione di Roma I,
      Universit\`a di Roma I, I-00185 Roma, ITALY}
        \\[-1mm]
  \hspace*{-4mm}
  {\small\it ${}^{\rm d}$
      Dept.~of Physics, New York University, 4 Washington Place,
      New York, NY 10003, USA}
        \\[4mm]
  {\small\tt Sergio.Caracciolo@mi.infn.it},\,
  {\small\tt tonyg@ms.unimelb.edu.au},\,
     \\[-2mm]
  \hspace*{-4mm}
    {\small\tt I.Jensen@ms.unimelb.edu.au},\,
    {\small\tt Andrea.Pelissetto@roma1.infn.it},  \\[-2mm]
  \hspace*{-4mm}
  {\small\tt anr@ms.unimelb.edu.au},\, {\small\tt sokal@nyu.edu}  \\[1cm]
}

\date{September 12, 2004 \\[1mm] revised May 25, 2005}

\maketitle
\thispagestyle{empty}   

\begin{abstract}
We study the correction-to-scaling exponents for the
two-dimensional self-avoiding walk,
using a combination of series-extrapolation and Monte Carlo methods.
We enumerate all self-avoiding walks up to 59 steps on the square lattice,
and up to 40 steps on the triangular lattice,
measuring the mean-square end-to-end distance, the
 mean-square radius of gyration and
the mean-square distance of a monomer from the endpoints.
The complete endpoint distribution
is also calculated for self-avoiding walks
 up to 32 steps (square) and up to 22 steps (triangular).
We also generate self-avoiding walks on the square lattice by Monte Carlo,
using the pivot algorithm, obtaining the mean-square radii
to $\approx 0.01\%$ accuracy up to $N = 4000$.
We give compelling evidence that the first non-analytic
correction term for two-dimensional self-avoiding walks is $\Delta_1 = 3/2$.
We compute several moments of the endpoint distribution function,
finding good agreement with the field-theoretic predictions.
Finally, we study a particular invariant ratio that can be shown,
by conformal-field-theory arguments, to vanish asymptotically,
and we find the cancellation of the leading analytic correction.
\end{abstract}

\vspace{1cm}

\newcommand{\be}{\begin{equation}}
\newcommand{\ee}{\end{equation}}
\newcommand{\<}{\langle}
\renewcommand{\>}{\rangle}
\def\bmath#1{\mbox{\protect\boldmath \protect\( #1 \protect\)}}
\def\reff#1{(\protect\ref{#1})}
\def\scrh{ {\cal H} }
\def\scrl{ {\cal L} }
\def\scro{ {\cal O} }
\def\scrs{ {\cal S} }
\def\spose#1{\hbox to 0pt{#1\hss}}
\def\ltapprox{\mathrel{\spose{\lower 3pt\hbox{$\mathchar"218$}}
 \raise 2.0pt\hbox{$\mathchar"13C$}}}
\def\gtapprox{\mathrel{\spose{\lower 3pt\hbox{$\mathchar"218$}}
 \raise 2.0pt\hbox{$\mathchar"13E$}}}

\newcommand{\zed}{{\bf \rm Z}}
\newcommand{\R}{\hbox{{\rm I}\kern-.2em\hbox{\rm R}}}

\def\half{{1 \over 2}}
\def\ehat{ {\hat{\bf e}} }
\def\myspace{ {\hspace{4.2mm}} }    
\def\myspaceb{ {\hspace{2.9mm}} }   

\clearpage

\section{Introduction}    \label{sec1}

The study of the universal properties of self-avoiding walks (SAWs)
in the long-chain limit has been a central problem in
both statistical mechanics and polymer physics for more than three decades
\cite{DeGennes_79,DesCloizeaux-Jannink_90,Schafer_99}.
In an $N$-step chain, the mean value of any global observable $\scro$
typically has an asymptotic expansion as $N\to\infty$ of the form
\begin{eqnarray}
   \<\scro\>_N   & = &
   A N^{p_\scro} \left[ 1 + {a_1 \over N} + {a_2 \over N^2} + \cdots +
                {b_0 \over N^{\Delta_1}} + {b_1 \over N^{\Delta_1 +1}}
                    + {b_2 \over N^{\Delta_1 +2}} + \cdots
         \right. \nonumber \\
   & & \qquad\qquad\qquad\left. +
                {c_0 \over N^{\Delta_2}} + {c_1 \over N^{\Delta_2 +1}}
                    + {c_2 \over N^{\Delta_2 +2}} + \cdots \right]
   \;,
\end{eqnarray}
where the leading exponent $p_\scro$ and the
correction-to-scaling exponents $\Delta_1 < \Delta_2 < \ldots$
are {\em universal}\/,
i.e.\ they depend on the spatial dimension $d$
but not on the microscopic details of the interactions
(provided that the interactions are short-ranged and primarily repulsive).
This universality justifies the intense efforts that have been
devoted to determining these universal exponents,
using a variety of analytical and numerical approaches.

In this paper we will address the problem of determining the
leading non-analytic correction-to-scaling exponent $\Delta_1$
for the two-dimensional self-avoiding walk
and for the closely related problem of self-avoiding polygons (SAPs).
At least two different theoretical predictions have been made
for the purportedly exact value of this exponent:
$\Delta_1 = 3/2$ based on Coulomb-gas arguments
\cite{Nienhuis_82,Nienhuis_84},
and $\Delta_1 = 11/16$ based on conformal-invariance methods
\cite{Saleur_87}.
In addition, several numerical methods have been employed
to estimate $\Delta_1$, notably
exact enumeration and extrapolation (series analysis)
\cite{Adler_83,Djordjevic_83,Majid_83,Privman_84,Guttmann_84,%
Rapaport_85,Ishinabe_88,Ishinabe_89,MacDonald_92,%
Conway-Enting-Guttmann,CG96,SCF96,JG99}
and Monte Carlo \cite{Havlin_83,Lyklema_85,Hunter_86,Lam_90,SCF96}.
The estimates of $\Delta_1$ resulting from these numerical works are,
for the most part, wildly contradictory:
even when one compares estimates produced by a single method,
such as series analysis, they range from
$\approx 0.5$ \cite{SCF96}
to $\approx 0.65$ \cite{Djordjevic_83,Privman_84,Ishinabe_88,%
Ishinabe_89}
to $\approx 0.85$ \cite{MacDonald_92}
to $\approx 1$ \cite{Adler_83,Ishinabe_89}
to 1.5 \cite{Conway-Enting-Guttmann,CG96,JG99}.
Similar variation can be found in estimates of $\Delta_1$
obtained from Monte Carlo studies, ranging from $\approx 0.5$~\cite{SCF96}
to $\approx 0.6$~\cite{Lam_90} to $\approx 0.84$~\cite{Lyklema_85}
to $\approx 1.1$ \cite{Lam_90}
to $\approx 1.2$ \cite{Havlin_83}.

Other models in the same universality class have also been considered,
yielding results in contrast with those for the SAW.
For instance, for lattice trails
(connected paths where the self-avoidance constraint
is applied only to bonds, not vertices)
it was shown by a transfer-matrix study \cite{GBB97}
that the correction-to-scaling exponent is indeed $\approx 11/16$,
confirming an earlier result based on series analysis \cite{CG93}.
This same transfer-matrix study also found $\Delta_1 \approx 3/2$
for SAWs. What remains to be understood is why the contribution
with $\Delta_1 = 11/16$ seems to be absent for square- and triangular-lattice
SAWs, yet present for trails.

To further confuse the subject, we should mention the
recent results of Jensen \cite{J02} for osculating SAPs
on the square lattice:
these are a superset of SAPs in which bonds may touch
at a vertex but not cross. In a sense they interpolate between SAPs,
in which intersections are strictly forbidden, and closed trails, in which
crossing at a vertex is allowed. A careful analysis of the corresponding
series \cite{J02} shows very convincingly that $\Delta_1 = 3/2$
and finds no correction corresponding to $\Delta_1 = 11/16$.

To add to our list of unexplained phenomena, we remark on a recent
Monte Carlo study of SAWs on the Manhattan lattice \cite{CCGP99},
where it was found that the critical exponent $\gamma$
has the same value 43/32 as for SAWs on regular lattices
{\em provided}\/ that the value $\Delta_1 = 11/16$ is used in the analysis.
It is quite unclear why a different value of $\Delta_1$
should arise for the Manhattan lattice than is found for the square lattice.

Returning now to the simplest case of SAWs and SAPs on regular lattices,
the gross disparities among the extant estimates
of the correction-to-scaling exponent might lead one to suspect
that different methods are computing different quantities.
For example, it might be that some methods are measuring (or predicting)
the leading correction exponent $\Delta_1$,
while others are measuring (or predicting) a subleading correction exponent
$\Delta_2$ or $\Delta_3$,
and still others are measuring some sort of ``effective''
exponent $\Delta_{\rm eff}$ that represents phenomenologically
the observed corrections to scaling in some specified interval
of walk length $N$
(and arising in reality from the sum of two or more correction-to-scaling
terms).

Moreover, it may even be true that different observables produce
different patterns of nonvanishing corrections to scaling.
For instance, the $\Delta_1 = 3/2$ correction term
appears to be present for SAWs but absent for SAPs.
While this may at first sight be considered
a violation of universality, we show below that it is not.

Two recent analyses \cite{CG96,JG99} based on very long series
for square-lattice SAWs and SAPs
have, however, yielded a consistent and convincing picture
of the corrections to scaling:
the first non-analytic correction-to-scaling exponent is indeed just
$\Delta_1 = 3/2$, as predicted by Nienhuis \cite{Nienhuis_82,Nienhuis_84};
but there are also analytic corrections to scaling
proportional to integer powers of $1/N$,
the first of which dominates asymptotically.
More precisely,
a careful numerical study based on a 51-term SAW series \cite{CG96}
found that the number of $N$-step SAWs on the square lattice is
given asymptotically by
\begin{eqnarray}
    c_N  & \sim &
    \mu^N N^{11/32} [1.177043 + 0.5500/N - 0.140/N^{3/2} - 0.12/N^2 + \cdots]
        \nonumber \\
    & & \;+\; (-\mu)^N N^{-3/2} [-0.1899 + 0.175/N - 1.51/N^2 + \cdots]
    \;.
 \label{eq1.1}
\end{eqnarray}
It is likely that previous studies identified some sort of effective exponent
that reflects a combination of the effects of the $1/N$ and $1/N^{3/2}$
correction-to-scaling terms
(see Section~\ref{sec2.2} for further discussion of this point).
Similarly, a careful numerical study based on a 90-term SAP series \cite{JG99}
found that the number of $2N$-step SAPs on the square lattice
is given asymptotically by
\be
    p_{2N} \;\sim\;
    \mu^{2N} N^{-5/2} [0.0994018 - 0.02751/N + 0.0255/N^2 + 0.12/N^3 + \cdots]
\;;
\ee
note that here there is {\em no}\/ $N^{-3/2}$ (or $N^{-5/2}$) correction term.
Finally, a recent transfer-matrix analysis \cite{GBB97} of SAWs
on the square lattice also found compelling numerical evidence in favour of 
the value $\Delta_1 = 3/2$, and against all values $\Delta_1$ significantly 
less than 3/2.

One possible cause of some confusion is that, because the value
of the leading critical exponent of the SAP generating function is
$2-\alpha = 3/2$,
any correction-to-scaling term with $\Delta=$ half-integer
``folds into'' the analytic background term and is therefore undetectable!
In other words, no corrections proportional to $N^{-\Delta}$ appear in the
coefficients $p_N$.
In order to understand this point, let us recall that
the critical exponent $\alpha$ is defined by the
leading asymptotic behavior $p_N \propto \mu^N N^{\alpha-3}$
of the polygon counts, corresponding to a leading behavior
$P(x) \equiv \sum_{N\ge 0} p_{N} x^N
 \sim {\rm const} \times (1-x/x_c)^{2-\alpha}$
as $x \uparrow x_c = 1/\mu$ for the polygon generating function.
If we now include both analytic and non-analytic corrections to scaling,
the polygon generating function can be written generically as
\be
   P(x) \;=\; \sum_{N\ge 0} p_{N} x^N \sim A(x) + B(x)(1-x/x_c)^{2-\alpha}
       [1 + c(1-x/x_c)^{\Delta_1} + \cdots ]
 \label{eq1.3}
\ee
with $A(x)$ and $B(x)$ analytic in the neighbourhood of
the critical point $x_c = 1/\mu$.
Since $\alpha = 1/2$ for two-dimensional SAPs
\cite{Nienhuis_82,Nienhuis_84,JG99},
if $\Delta_1 =$ half-integer the above equation may be rewritten as
\be
   P(x) \;=\;  \sum_{N\ge 0} p_{2N} x^N
       \;\sim\;  \widehat{A}(x) +\widehat{B}(x)(1-x/x_c)^{3/2} [1 + \cdots]
   \;,
\ee
with the $(1-x/x_c)^{\Delta_1}$ correction term absorbed into
the analytic background term $\widehat{A}(x)$.
Therefore, if \reff{eq1.3} holds, no correction
of the form $N^{-\Delta_1}$ is present.
On the other hand, if the polygon counts were to exhibit a behaviour of the
form
$p_N \propto \mu^N N^{\alpha-3} [1 + \ldots + a/N^{\Delta_1} + \ldots]$
with $\alpha = 1/2$ and $\Delta_1 = 3/2$
--- and hence include a term $\propto \mu^N N^{-4}$ ---
then the generating function $P(x)$ would exhibit,
on top of the $(1-x/x_c)^{3/2}$ leading behaviour,
a non-analytic confluent term of the form $(1-x/x_c)^3 \log(1-x/x_c)$
in addition to the analytic term $(1-x/x_c)^3$.
However, as discussed in some detail in \cite{JG99},
there is no evidence for a term of the form $a/N^{3/2}$
in the analysis of the SAP count series,
and indeed there is considerable evidence for the absence of such a term.
There is, however, abundant evidence of such a term in the 
radius-of-gyration series of SAPs \cite{JG00}.

In the present paper we make some further progress
in supporting the assertion that $\Delta_1 = 3/2$
for SAWs on regular two-dimensional lattices (here square and triangular).
First, we make a conventional analysis of corrections to scaling
in the standard observables $\< R_e^2 \>$, $\< R_g^2 \>$ and $\< R_m^2 \>$;
our contribution here is to present and use extended series expansions
and a more efficient Monte Carlo algorithm.
The results of this analysis are consistent with other recent work in
supporting the conclusion that $\Delta_1 = 3/2$.
In the course of this analysis we point out that, for certain observables,
pairs of correction terms of opposite sign can (and do) conspire
to give an effective exponent that is smaller than both of the
individual exponents;
this explains the apparent exponents $\Delta < 1$ observed in some
earlier work.
Second---and this is perhaps our main contribution---we point out
several observables in which
{\em a correction-to-scaling term becomes the leading term}\/.
These include:
(a) the combination
${246 \over 91} \< R_g^2 \> - 2 \< R_m^2 \> + {1 \over 2} \< R_e^2 \>$,
which arises in the conformal-invariance theory
\cite{Cardy-Saleur,CPS:Cardy-Saleur};
and (b) quantities related to the breaking of Euclidean invariance
down to the lattice symmetry group, the simplest of which is
(on the square lattice) the  fourth-order moment
$\< r^4 \cos 4\theta \> = \< x^4 - 6x^2 y^2 + y^4 \>$.
Analysis of these quantities by Monte Carlo methods yields only a modest
improvement over the analysis of conventional quantities---the
trouble is that the new quantities exhibit a low
``signal-to-noise ratio''---but the series analysis is quite precise.

The plan of this paper is as follows.
In Section~\ref{sec2} we define the quantities to be studied and collect
some theoretical results that will be used or tested in the following sections.
Section~\ref{sec3} reports the results of our series analysis:
first, we analyze the SAW counts (Section~\ref{sec3.3});
then, we analyze the radius of gyration,
the end-to-end distance and the average distance of a monomer from the
endpoints, along with their invariant ratios (Section~\ref{sec3.4});
finally, we analyze the higher-order rotationally-invariant moments
of the endpoint distribution function (Section~\ref{sec3.5})
and the corresponding non-rotationally-invariant moments
(Section~\ref{sec3.6}).
For each of them, we determine the asymptotic behaviour as $N\to\infty$,
focusing in particular on the correction-to-scaling exponent
$\Delta_1$ and on the behaviour at the antiferromagnetic singularity 
(in the case of the square lattice).
In Section~\ref{sec4} we report the analyses of our Monte Carlo data,
confirming the absence of a correction-to-scaling exponent $\Delta_1 = 11/16$.
Finally, in Section~\ref{sec5} we draw our conclusions.


\section{Definitions and theoretical background}    \label{sec2}

In this section we review briefly the basic facts and conjectures
about the SAW that will be used (or tested) in the remainder of the paper.

\subsection{Definitions and notation}   \label{sec2.1}

Let $\scrl$ be some regular $d$-dimensional lattice.
Then an {\em N-step self-avoiding walk}\/ (SAW) $\omega$ on $\scrl$
is a sequence of {\em distinct}\/ points
$\omega_0 ,\omega_1 ,...,\omega_N$ in $\scrl$
such that each point is a nearest neighbour of its predecessor.
We assume all walks to begin
at the origin ($\omega_0 = 0$) unless stated otherwise.

First we define the quantities relating to the {\em number}\/ (or ``entropy'')
of SAWs.
Let $c_N$ [resp.\ $c_N({\bf x})$] be the number of $N$-step SAWs on $\scrl$
starting at the origin and ending anywhere [resp.\ ending at ${\bf x}$].
Then $c_N$ and $c_N({\bf x})$ are believed to have the asymptotic behaviour
\begin{eqnarray}
  c_N      & \sim &   {\rm const} \times \mu^N  N^{{\gamma} -1}
      \label{gamma} \\
  c_N ({\bf x})  & \sim &   {\rm const} \times \mu^N  N^{{\alpha} -2}
                           \qquad ({\bf x} {\rm\ fixed} \ne 0)
\label{alpha}
\end{eqnarray}
as $N\to\infty$;  here $\mu$ is called the {\rm connective constant}\/
of the lattice, and $\gamma$ and $\alpha$ are {\em critical exponents}\/.
The critical exponents are believed to be universal among lattices of a
given dimension $d$.  For rigorous results concerning the asymptotic behaviour
of $c_N$ and $c_N({\bf x})$, see
\cite{Madras-Slade,Hara-Slade_1,Hara-Slade_2,Hara-Slade-Sokal}.

Next we define several measures of the {\em size}\/ of an $N$-step SAW:
\begin{itemize}
  \item  The {\em squared end-to-end distance}\/
\end{itemize}
\be
  R_e^2    \;=\;   \omega_N^2   \;.
\ee
\begin{itemize}
  \item  The {\em squared radius of gyration}\/
\end{itemize}
\be
  R_g^2   \;=\;
        {1 \over 2(N+1)^2}  \sum_{i,j=0}^N (\omega_i - \omega_j)^2 \;.
\ee
\begin{itemize}
  \item  The {\em mean-square distance of a monomer from the endpoints}\/
\end{itemize}
\be
  R_m^2   \;=\;  {1 \over 2(N+1)}  \sum_{i=0}^N
        \left[ \omega_i^2  \,+\,  (\omega_i-\omega_N)^2  \right]   \;.
\ee
We then consider the mean values $\< R_e^2 \>_N$, $\< R_g^2 \>_N$ and
$\< R_m^2 \>_N$ in the probability distribution that gives equal weight
to each $N$-step SAW.
Very little has been proven rigorously about these mean values,
but they are believed to have the leading asymptotic behaviour

\be
  \< R_e^2 \> _N  ,\,
  \< R_g^2 \> _N  ,\,
  \< R_m^2 \> _N       \;\sim\;  {\rm const} \times N^{{2} \nu}   \label{nu}
\ee
as $N \to \infty$, where $\nu$ is another (universal) critical exponent.
Hyperscaling \cite{PV02} predicts that
\be
   d\nu  \;=\;  2-\alpha  \;.
\ee

For SAWs in two dimensions,
Coulomb-gas arguments \cite{Nienhuis_82,Nienhuis_84}
as well as arguments based on stochastic Loewner evolution (SLE)
\cite{Lawler_02}
predict that $\nu = 3/4$, $\alpha = 1/2$ and $\gamma=43/32$.
Prior numerical studies have confirmed these values to
high precision \cite{Li-Madras-Sokal,CG96,JG99};
in this paper we take them for granted.

The amplitude ratios
\begin{eqnarray}\label{ratios}
     A_N   & = &   {\< R_g^2 \> _N   \over   \< R_e^2 \> _N}       \\[1mm]
     B_N   & = &   {\< R_m^2 \> _N   \over   \< R_e^2 \> _N}       
\end{eqnarray}
are expected to approach universal values in the limit $N \to \infty$,
which we call $A$ and $B$;
one of our goals is to estimate these limiting amplitude ratios.
Many other universal amplitude combinations (notably involving SAPs)
are discussed in \cite{Cardy-Guttmann,Cardy-Mussardo}.

Of particular interest is the linear combination
\cite{Cardy-Saleur,CPS:Cardy-Saleur}
\be\label{CSC}
 F_N \;\equiv\;
   \left( 2 + {\displaystyle y_t \over \displaystyle y_h} \right)  A_N
   \,-\, 2 B_N  \,+\, {1 \over 2}
\ee
and the corresponding unnormalized quantity
\be
 f_N  \;\equiv\;  F_N \< R^2_e\>_N   \;\equiv\;
   \left( 2 + {\displaystyle y_t \over \displaystyle y_h} \right) \< R^2_g
\>_N
   \,-\, 2 \< R^2_m \>_N  \,+\, {1 \over 2} \< R^2_e\>_N
 \;,
\label{fN-def}
\ee
where $y_t = 1/\nu$ and $y_h = 1 + \gamma/(2\nu)$
are the thermal and magnetic renormalization-group eigenvalues, respectively,
of the $n$-vector model at $n=0$.
In two dimensions
---where $y_t = 4/3$ and $y_h = 91/48$, hence $2 + y_t/y_h = 246/91$---Cardy
and Saleur \cite{Cardy-Saleur}
(as corrected by Caracciolo, Pelissetto and Sokal \cite{CPS:Cardy-Saleur})
have predicted, using conformal field theory,
that $\lim_{N\to\infty} F_N = 0$.
We shall henceforth refer to this relation as the CSCPS relation.
This conclusion has been confirmed by previous
high-precision Monte Carlo work \cite{CPS:Cardy-Saleur}
as well as by series extrapolations \cite{Guttmann-Yang}.
It is therefore of interest to examine the {\em rate}\/
at which $F_N$ tends to zero, as this gives information
on the correction-to-scaling terms.
We will discuss this from a theoretical point of view
near the end of Section~\ref{sec2.2},
and from a numerical point of view in Sections~\ref{sec3.5.2} and \ref{sec4.2}.

It turns out that $\lim_{N\to\infty} F_N = 0$ holds not only
for the ordinary square-lattice SAW,
but also for SAWs with nearest-neighbour interactions,
right up to (but not at) the theta point \cite{OPBG}.
Moreover, the relation appears to hold {\em at}\/ the theta point
if $2 + y_t/y_h$ is given its theta-point value 23/8 instead of 246/91.
This observation is used in~\cite{OPBG} to locate the theta point
more precisely.

We shall also consider higher moments of the end-to-end distance.
Limiting ourselves to two-dimensional lattices, let us write
\be
   \omega_N   \;\equiv\;   (x,y)   \;\equiv\;   (r\cos\theta,r\sin\theta)  \;.
\ee
The Euclidean-invariant moments $\< r^k \>_N$
are of course expected to behave as
\be
   \< r^k \>_N   \;\sim\;  {\rm const} \times N^{k\nu}
\ee
as $N \to \infty$. One can consider the dimensionless ratios
\be
  M_{2k,N} \;=\; {\< r^{2k} \>_N\over \< r^2 \>_N^k}   \;,
\ee
which approach finite limits for $N\to\infty$;
these limiting ratios $M_{2k,\infty}$ are universal quantities
that characterize the end-to-end distribution function.
Estimates of $M_{2k,\infty}$ have been obtained in
\cite{Caracciolo-etal-distributionf}
using field theory and the Laplace--deGennes transform method.
It turns out \cite{CPRV-2punti,CPRV-2dNle2}
that the 2-point function is very nearly equal to
that of a free field,
so that when the rescaled inverse propagator in momentum space\footnote{
   $\widetilde{D}(q)$ is the Fourier transform of the two-point
   correlation function, rescaled so that the first two terms
   at small $q$ are $1-q^2 + O(q^4)$.
}
is written as
\be
   \widetilde{D}(q)^{-1}   \;=\;  1 + q^2 + \sum_{n=2}^\infty b_n q^{2n}
   \;,
\ee
one has $1 \gg |b_2| \gg |b_3| \gg |b_4| \gg \ldots\;$.
One obtains
\cite{Caracciolo-etal-distributionf}
\be
M_{2k,\infty}  \;=\;
    {\Gamma(\gamma + 2 \nu)^k
     \over
     \Gamma(\gamma + 2 k \nu) \, \Gamma(\gamma)^{k-1}
    }
    \, \left[1 - b_2 (k-1) + R_k\right] \, k! \,
    \prod_{j=0}^{k-1} \left(1 + {2 j\over d}\right)   \;,
 \label{momenti-theory}
\ee
where $R_k$ is a very small correction (unless $k$ is very large)
that involves the
constants $b_3$, $b_4$, $\ldots$ as well as higher powers
$b_i b_j, b_i b_j b_k, \ldots\;$.
Explicitly,
\be
R_k  \;=\;  \sum_{n=3}^{k} (-1)^{n+1}(k+1-n) b_n  \,+\,
      {1\over2} (k-2)(k-3) b_2^2  \,+\, \cdots   
\ee
Note that $R_2 = 0$ exactly.
The universal nonperturbative constants $b_2,b_3,\ldots$ have been obtained
from the analysis of exact-enumeration series
on the square, triangular and hexagonal lattices \cite{CPRV-2dNle2}.
Numerically, it is found \cite{CPRV-2dNle2}
that $b_2$ is extremely small, $b_2 = 0.00015(20)$,
and that $b_3$ is even smaller, $|b_3| \ltapprox 3 \times 10^{-5}$.
Using the estimate of $b_2$ in \reff{momenti-theory} and 
neglecting $R_k$, we obtain for the lowest values of $k$:
\begin{eqnarray}
M_{4,\infty} &=& 1.44574(28) \label{M4CPRV}\\
M_{6,\infty} &=& 2.5876(10)  \label{M6CPRV}\\
M_{8,\infty} &=& 5.3805(32)  \label{M8CPRV}\\
M_{10,\infty} &=& 12.557(10)\; . \label{M10CPRV}
\end{eqnarray}

A second class of interesting observables are moments
that are invariant under the symmetry group of the lattice
but {\em not}\/ under the full Euclidean group:
examples are the moments $\< r^k \cos n\theta \>$ with $n \neq 0$,
where for the square (resp.\ triangular) lattice
$n$ must be a multiple of 4 (resp.\ 6).
We expect these non-Euclidean-invariant moments to behave as
\be
   \< r^k \cos n\theta \>   \;\sim\;
   {\rm const} \times   N^{k\nu - \Delta_{\rm nr}}  \;,
\ee
where $\Delta_{\rm nr} > 0$ is a new correction-to-scaling exponent
\cite{CPRV-2punti}
associated with the breaking of full rotation invariance
down to the lattice rotation group:
it thus depends on the lattice in question (e.g., square or triangular)
and is in general different from the leading correction-to-scaling exponent
$\Delta_1$ (which corresponds to a Euclidean-invariant irrelevant operator).

For Gaussian models---and thus also for $n$-vector models
(including the SAW case $n=0$) in dimension $d \ge 4$---we
have $\Delta_{\rm nr} = 2\nu=1$ on any hypercubic lattice.
For $n$-vector models in dimension $d=4-\epsilon$,
this relation is modified at order $\epsilon^2$ \cite{CPRV-2punti}:
\be
    \Delta_{\rm nr} \;=\; \nu\left[
    2 \,+\, {7 \over 20} {n+2 \over (n+8)^2} \epsilon^2 \,+\, O(\epsilon^3)
     \right]
    \;.
 \ee
In dimension $d=3$,
several alternative methods---field theory and exact-enumeration
analysis---show that $\Delta_{\rm nr}$
is very close to $2\nu$, though not exactly equal \cite{CPRV-2punti}.
In two dimensions on the square lattice, $\Delta_{\rm nr} = 2\nu$ exactly
for the Ising model and for the $n$-vector model with $n\ge 3$
(in the latter case with logarithmic corrections) \cite{CPRV-2punti}.
For the triangular lattice,
similar arguments \cite{CPRV-2punti} predict
$\Delta_{\rm nr} = 4 \nu$.\footnote{
   For the hexagonal lattice, $\Delta_{\rm nr} = 4 \nu$ 
   for observables that break rotational invariance but
   are invariant under interchange of the two sublattices,
   while $\Delta_{\rm nr} = 3 \nu$ for observables
   that distinguish the two sublattices \cite{CPRV-2punti}.
}
For the Ising model, these predictions can be obtained using
conformal field theory (see \cite{Caselle-Hasenbusch_99,CHPV_01}
for the classification of the subleading operators appearing in the
Ising model);
they can be checked explicitly \cite{CPRV-2punti}
for at least one specific observable,
using the analytic expression for the mass gap
\cite{Cheng-Wu_67,Stephenson,CPRV-2dNle2}.\footnote{
   For instance, consider on a square lattice
   the mass gap $m(\hat{n})$ in the direction $\hat{n}$ defined as
   $$
      m(\hat{n})  \;=\;  - \lim_{r\to\infty} {1\over |r|}
          \log \left( \sum_{\vec{x}\cdot \hat{n} = r} G(\vec{x})\right)
      \;,
   $$
   were $\hat{n} = (\cos\theta,\sin\theta)$ is a unit vector
   and the summation runs over all $\vec{x}$ such that
   $\vec{x}\cdot \hat{n} = r$.
   From the exact solution \cite{Cheng-Wu_67}, one can easily see that,
   for $\beta\to\beta_c$,
   $$
      m(\hat{n}) \;=\; m_0 (\beta_c - \beta)^{-1} \left[1 +
         (\beta_c - \beta)^2 (a_0 + b_0 \cos 4 \theta) + \cdots \right]
   $$
   with $b_0 \neq 0$.
   This result shows explicitly that $\Delta_{\rm nr} = 2 = 2\nu$.
}
It is therefore suggestive to conjecture that the same relations
between $\Delta_{\rm nr}$ and $\nu$ are valid for the SAW.
This would predict $\Delta_{\rm nr} = 3/2$ on the square lattice,
and $\Delta_{\rm nr} = 3$ on the triangular lattice.
In Section~\ref{sec3.6} we will test (and confirm) this conjecture,
by series analysis, for both square-lattice and triangular-lattice SAWs.

\subsection{Corrections to scaling}   \label{sec2.2}

Let us now make some general remarks concerning corrections to scaling.
Clearly, \reff{gamma}/\reff{alpha}/\reff{nu}
are only the leading term in a large-$N$ asymptotic expansion.
According to renormalization-group theory \cite{Wegner_72},
the mean value of any global observable $\scro$ behaves as $N\to\infty$ as
\begin{eqnarray}
   \<\scro\>_N   & = &
   A N^{p_\scro} \left[ 1 + {a_1 \over N} + {a_2 \over N^2} + \cdots +
                {b_0 \over N^{\Delta_1}} + {b_1 \over N^{\Delta_1 +1}}
                    + {b_2 \over N^{\Delta_1 +2}} + \cdots
         \right. \nonumber \\
   & & \qquad\qquad\qquad\left. +
                {c_0 \over N^{\Delta_2}} + {c_1 \over N^{\Delta_2 +1}}
                    + {c_2 \over N^{\Delta_2 +2}} + \cdots \right]
   \;.
 \label{eq_wegner}
\end{eqnarray}
Thus, in addition to ``analytic'' corrections to scaling
of the form $a_k/N^k$,
there are ``non-analytic'' corrections to scaling of the form
$b_k/N^{\Delta_1 +k}$, $c_k/N^{\Delta_2 +k}$ and so forth,
as well as more complicated terms [not shown in \reff{eq_wegner}]
which have the general form
${\rm const}/N^{k_1 \Delta_1 + k_2 \Delta_2 + \cdots + l}$
where $k_1,k_2,\ldots$ and $l$ are non-negative integers.
The leading exponent $p_\scro$ and the correction-to-scaling exponents
 $\Delta_1 < \Delta_2 < \ldots$ are universal;
$p_\scro$ of course depends on the observable $\scro$ in question,
 but the $\Delta_i$ do not.
The various amplitudes (both leading and subleading) are all nonuniversal
(and of course also depend on the observable\footnote{
   Sometimes a particular correction-to-scaling amplitude
   will vanish for some observables but not for others
   (e.g.\ for symmetry reasons).
}).
However, {\em ratios}\/ of the corresponding amplitudes $A$, $b_0$ and $c_0$
(but not $a_k$ or the higher $b_k, c_k$)
for different observables are universal \cite{Privman_91,Nickel_91}.

In fact, \reff{eq_wegner} is incomplete,
as there are ``mixing'' terms arising from the fact that
the temperature deviation from criticality is a smooth but {\em nonlinear}\/
function of the nonlinear scaling fields $g_t$ and $g_h$.
This has the consequence
\cite{Aharony_83,Gartenhaus_87,Gartenhaus_88,Salas-Sokal_9904038v1,PV02}
that the susceptibility (or SAW generating function),
which has a leading singularity $(x_c - x)^{-\gamma}$,
also contains an additive term proportional to the energy,
of order $(x_c - x)^{1 - \alpha}$.
In the case of the two-dimensional Ising model, we have $\alpha = 0$,
and this term is responsible for the logarithmic terms
in the susceptibility, as was recently exhaustively studied in~\cite{ONGP01}.
For the two-dimensional SAW, we have $\alpha = 1/2$,
and so one would expect a term
$\widetilde{A}(x) (x_c - x)^{1/2}$ in the SAW generating function.
To incorporate this term requires that the naively expected
asymptotic form
\begin{equation}
c_N  \;\sim\;
    \mu^N N^{11/32}[a_0 + a_1/N + a_2/N^{3/2} + a_3/N^2 + a_4/N^{5/2} + \cdots]
\end{equation}
be modified to read
\begin{eqnarray}
c_N & \sim &
    \mu^N N^{11/32}[a_0 + a_1/N + a_2/N^{3/2} + 
          a_3/N^2 + a_4/N^{5/2} + \cdots]  \nonumber \\
   & & \quad + \; \mu^N N^{-3/2} [\widetilde{a}_0 + \widetilde{a}_1/N + \cdots]
    \;.
 \label{eq.CN.modified}
\end{eqnarray}

For loose-packed (i.e., bipartite) lattices,
such as the square and simple-cubic lattices,
there is an additional set of terms
arising from the antiferromagnetic singularity,
of the form
\begin{equation}
   (-1)^N N^q \left[ d_0 + {d_1 \over N} + {d_2 \over N^2} + \ldots +
                {e_0 \over N^{\Delta_1^{\rm AF}}} +
                {e_1 \over N^{\Delta_1^{\rm AF} +1}} +
                {e_2 \over N^{\Delta_1^{\rm AF} +2}} + \ldots \right]
   \;,
 \label{corr_AF}
\end{equation}
where the exponent $q$ of course depends on the observable.
We know of no theoretical argument that
predicts the value of the exponent $\Delta_1^{\rm AF}$.
For the exponent $q$, in the closely related problem of the Ising-model
susceptibility
in two and three dimensions, Sykes~\cite{SF58} has given a configurational
``counting theorem'' that enables one to guess that the antiferromagnetic
susceptibility behaves as the internal energy. This reasoning is discussed in
greater detail in \cite{Fisher_62,MEF65}.\footnote{
   The basic idea is that the susceptibility can be rewritten
   as the sum of two terms:
   one proportional to the energy,
   and a second one which can be argued (by series analysis)
   to give an algebraically small contribution near the antiferromagnetic critical point.
}
It follows that there should be a term in the susceptibility
of the form $D(x)(1 + x/x_c)^{1-\alpha}$
[where $D(x)$ is analytic in a neighbourhood of the
antiferromagnetic critical point $x=-x_c$]
and thus a term $(-x_c)^{-N} N^{\alpha-2}$ in the
high-temperature-series coefficients.
This result can be put on more solid ground \cite{BG-93} by noting
that at the antiferromagnetic critical point the (unstaggered) magnetic field
is an irrelevant variable, so that the leading contribution
to the free energy is
\be
F(x,h) \;=\; a g_{t}(x,h)^{2-\alpha} + F_{\rm reg}(x,h),
\ee
where $x$ is the inverse temperature
and $g_{t}(x,h)$ is the nonlinear scaling field
associated with the temperature at the antiferromagnetic critical point.
Since
\be
g_{t}(x,h) \;=\; (1 + x/x_c) +a_t h^2 + \ldots   \;,
\ee
by performing the appropriate derivatives
we obtain the result reported above
(provided of course that $a_t \neq 0$).
This argument is very general and applies to any $n$-vector model;
in particular, it applies for $n=0$, i.e.\ to the SAW.
Thus, for the SAW counts $c_N$ we expect a term $(-\mu)^N N^{\alpha-2}$,
so that $q = \alpha-2$ for this observable \cite{GW-78}.

The argument of Sykes can be generalized to higher moments of the two-point 
function, i.e., $\sum_r |r|^{2k} G(r)$. Also in this case one can
identify two terms: one is proportional to the energy, while the other 
is conjectured to give an algebraically small (i.e., noncritical) correction 
at the antiferromagnetic critical point. 
Such a conjecture was numerically verified in \cite{CPRV-02} for the 
three-dimensional Ising model. As for the susceptibility, this implies
that asymptotically the moments have the form $D(x)(1 + x/x_c)^{1-\alpha}$,
with $D(x)$ analytic, for any $k$.
Therefore, a term $(-\mu)^{N} N^{\alpha-2}$ should be present in their
high-temperature-series coefficients. 
Extending this conjecture to the 
$n$-vector model and in particular to the SAW ($n=0$), we predict
\begin{eqnarray}
c_N  &\sim&  \mu^N N^{\gamma-1}[a_0 + \cdots]
         \;+\; (-\mu)^N N^{\alpha-2}[d_0 + \cdots]
     \;,
  \label{q1}
\\
c_N \langle r^{2k}\rangle_N &\sim& 
\mu^N N^{2k\nu+\gamma-1}[a'_0 + \cdots]
         \;+\; (-\mu)^N N^{\alpha-2}[d'_0 + \cdots]
     \;.
  \label{q2} 
\end{eqnarray}
For $k=1$, \reff{q2} gives the behaviour of $c_N \langle R^2_e \rangle_N$.
It may seem natural to generalize the expression \reff{q2} to the other
metric quantities, namely $R^2_m$ and $R^2_g$. 
Surprisingly (to us), our subsequent analysis (see Section~\ref{sec3.5.2})
shows that, while the unnormalized second-moment series
of the end-to-end distance series behaves precisely as
expected in \reff{q2}, the unnormalized series corresponding to both
the radius of gyration and the mean monomer-endpoint distance behave a
little differently. We find
\be
c_N \<R_{g,m}^2\>_N  \;\sim\;  \mu^N N^{2\nu+\gamma-1}[a'_0 + \cdots]
         \;+\; (-\mu)^N [d'_0 + \cdots]
     \;.
  \label{q2b}
\ee
That is to say, the antiferromagnetic exponent is different in the
latter cases, namely $0$ instead of $\alpha-2 = -3/2$.
Nevertheless, by taking the quotient of either \reff{q2} or \reff{q2b}
by \reff{q1}, we obtain in all cases
\be
\<R^2\>_N  \;\sim\;  N^{2\nu}[a''_0 + \cdots]
         \;+\; (-1)^N N^q [d''_0 + \cdots]
  \label{q2a}
\ee
with $q = 2\nu + \alpha - 1 - \gamma$.
[For the end-to-end distance, the {\em dominant}\/ antiferromagnetic
correction always comes only from \reff{q1};
for the other two metric quantities, it comes from both \reff{q1} and
\reff{q2b}.]
For the end-to-end distance only, we have the additional relation
\be
{a''_0\over d''_0}  \;=\;  - {a_0\over d_0}  \;.
\label{rel-a0d0}
\ee

Similarly, the rotationally-invariant higher moments $\< r^{2k} \>_N$
are expected to behave as
\be
\< r^{2k} \>_N  \;\sim\;  N^{2k\nu}[a'''_0 + \cdots]
         \;+\; (-1)^N N^{q_k} [d'''_0 + \cdots]
  \label{q2k}
\ee
with $q_k = 2k\nu + \alpha - 1 - \gamma$. The coefficients $a'''_0$ and 
$d'''_0$ also satisfy a relation analogous to \reff{rel-a0d0}.
Our numerical analysis, described below,
provides supporting evidence that the corresponding exponents
are indeed $q_k = 3k/2 - 59/32$ in two dimensions
(see Sections~\ref{sec3.5.2} and \ref{sec3.5}).

Finally, the non-analytic correction-to-scaling exponent
$\Delta_1^{\rm AF}$ was found numerically,
in the case of the square-lattice SAW counts, to be $1$
\cite{CG96}.
It would seem likely that this value should also hold for other properties,
such as the metric quantities $\<R^2\>_N$.
Our numerical studies, discussed below, are consistent with this 
conjecture---or, put another way, they are insufficiently sensitive
to refute this obvious first guess.

Let us now return to the question of the corrections to the
CSCPS relation $\lim_{N\to\infty} F_N = 0$ [cf.\ \reff{CSC}].
Series analysis and Monte Carlo simulations
(see Sections~\ref{sec3.5.2} and \ref{sec4.2} below)
indicate that $F_N \propto N^{-3/2}$,
i.e.\ that the leading analytic correction cancels.
Such a cancellation may seem surprising, but it can be understood
by means of a standard renormalization-group argument.
Consider the continuum $O(n)$ model with Hamiltonian
\be
{\cal H} \;=\; {\cal H}^* \,+\, \int d^2r\, [t E(r) + h s^1(r)]  \;,
\ee
where ${\cal H}^*$ is the fixed-point Hamiltonian,
and $E(r)$ and $s^i(r)$ are the energy and spin operators, respectively.
The CSCPS relation \cite{Cardy-Saleur,CPS:Cardy-Saleur}
is a consequence of the sum rule
\be
\int d^2r \, \langle \Theta(0)^{\rm cont} \Theta(r)^{\rm cont} \rangle
  \;=\; 0  \;,
\ee
where $\Theta(r)^{\rm cont}$ is the trace of the
continuum stress-energy tensor,
and of course we must set $n = 0$ to obtain SAWs.
In order to translate this continuum relation into a lattice one,
we must relate the continuum operator to its lattice counterpart.
It is natural to assume that
the trace of the lattice stress-energy tensor, $\Theta(r)^{\rm latt}$,
whose explicit form is given in \cite{Cardy-Saleur},
behaves as
\be
\Theta(r)^{\rm latt}  \;=\;  Z(t,h) \Theta(r)^{\rm cont} \,+\, \ldots  \;,
\label{theta-latt-cont}
\ee
where $Z(t,h)$ is a smooth function of $t$ and $h$, and
the dots represent the contributions of the subleading operators.
As a consequence of (\ref{theta-latt-cont}) we have
\be
\int d^2r \, \langle \Theta(0)^{\rm latt} \Theta(r)^{\rm latt} \rangle 
   \;=\; {\rm O}(t^{\Delta_1}, h^{(y_t/y_h)\Delta_1})  \;.
\label{latt-ThTh}
\ee
No corrections of order $t$ appear in the previous relation.
Eq.~(\ref{latt-ThTh}) therefore implies
the absence of the analytic corrections
in the CSCPS relation $\lim_{N\to\infty} F_N = 0$.

The observation that $F_N \propto N^{-3/2}$
implies a constraint on the subdominant amplitudes.
More precisely, if we write
\begin{eqnarray}\label{27}
\<R_e^2\>_N &\sim & a_e N^{3/2} + b_e N^{1/2} + c_e + {\rm O}(1/\sqrt N)
    \label{27a} \\
\<R_g^2\>_N &\sim & a_g N^{3/2} + b_g N^{1/2} + c_g + {\rm O}(1/\sqrt N)
    \label{27b} \\
\<R_m^2\>_N &\sim & a_m N^{3/2} + b_m N^{1/2} + c_m + {\rm O}(1/\sqrt N),
    \label{27c}
\end{eqnarray}
then the original CSCPS relation $F_N \to 0$ implies
\be \label{leading}
   91 a_e  \;=\; 364 a_m - 492 a_g  \;,
\ee
while the absence of a $1/N$ term in $F_N$
means that the leading subdominant terms
also satisfy an amplitude relationship analogous to \reff{leading},
namely
\be
   91 b_e  \;=\; 364 b_m - 492 b_g  \;.
 \label{eq_CSCPS_subdominant}
\ee
Note too from \reff{fN-def} that
\be
   f  \;\equiv\; \lim_{N \rightarrow \infty} f_N
      \;=\; (2 + y_T/y_H) c_e - 2 c_m + c_g/2.  \;
 \label{eq_CSCPS_f}
\ee

Let us conclude by discussing briefly the behaviour
of ``effective exponents''.
Given a function $f(N)$, let us define $\Delta_{\rm eff}(N)$
by fitting locally to the Ansatz $f(N) = a + b/N^\Delta$:
this gives
\be
   \Delta_{\rm eff}(N)  \;\equiv\;
   -{d \log f'(N) \over d  \log N} \,-\, 1 \;.
\ee
Applying this to $f(N) = a_0 + a_1/N^{\Delta_1} + a_2/N^{\Delta_2}$,
we obtain
\be
   \Delta_{\rm eff}(N)  \;=\;
    {a_1 \Delta_1^2 N^{-\Delta_1} + a_2 \Delta_2^2 N^{-\Delta_2}
     \over
     a_1 \Delta_1 N^{-\Delta_1} + a_2 \Delta_2 N^{-\Delta_2}
    } \;.
\ee
Thus, if $a_1$ and $a_2$ have the same sign,
the ``effective'' exponent $\Delta_{\rm eff}(N)$
lies between $\Delta_1$ and $\Delta_2$ for all $N$,
and decreases monotonically to $\Delta_1$ as $N \to\infty$.
(This is the behaviour one would expect intuitively
 for an ``effective exponent''.)
By contrast, if $a_1$ and $a_2$ have opposite signs,
then $\Delta_{\rm eff}(N)$ starts {\em above}\/ $\Delta_2$ for small $N$,
{\em increases}\/ monotonically and reaches $+\infty$
at some finite value of $N$;
it then jumps to $-\infty$,
after which it continues to increase monotonically,
tending asymptotically to $\Delta_1$ {\em from below}\/ as $N \to\infty$.
Thus, the qualitative behaviour of the effective exponents
depends crucially on the sign (and magnitude) of $a_1/a_2$,
which can vary from one observable to another.
We shall see this phenomenon quite clearly in the two-dimensional SAW.

\section{Series analysis}   \label{sec3}

\subsection{Summary of our data}  \label{sec3.1}


We have previously reported enumerations of square-lattice SAWs
up to 29 steps for $c_N$, $\<R_e^2\>_N$, $\<R_g^2\>_N$ and $\<R_m^2\>_N$
\cite{Guttmann-Wang_91,BW98}, and of triangular-lattice SAWs
up to 22 steps for $c_N$ and $\<R_e^2\>_N$ \cite{Guttmann-Wang_91}
and up to 19 steps for $\<R_g^2\>_N$ and $\<R_m^2\>_N$ \cite{BW98}.
Analysis of these series can be found in \cite{Guttmann-Wang_91,BW98}.
We have also previously presented the square-lattice SAW counts $c_N$
up to 51 steps \cite{CG96}
and the square-lattice polygon counts $p_{2N}$
up to 90 steps \cite{JG99}.
Analysis of these SAW series \cite{CG96,JG99}
provided good evidence that the
non-analytic correction-to-scaling exponent
is exactly $\Delta_1 = 3/2$ as predicted by Nienhuis
\cite{Nienhuis_82,Nienhuis_84},
and that there is also the expected analytic term of leading order $1/N$
(as well as $1/N^2, 1/N^3, \ldots$). For SAPs we found compelling evidence 
for purely analytic correction-to-scaling terms.
We have thus far found no numerical evidence
of a second non-analytic correction-to-scaling exponent $\Delta_2$,
although it is reasonable to expect that one exists.

In the present paper, we have extended the previous work by
enumerating all SAWs on the square lattice up to 59 steps,
and on the triangular lattice up to 40 steps, using
refinements of the finite-lattice method (FLM)
due to Rogers (unpublished) and Jensen~\cite{J04a}.
The results for $c_N$, $\< R_e^2 \> _N$, $\< R_g^2 \> _N$ and $\< R_m^2 \> _N$
are collected in Tables \ref{table1} and \ref{table2}.

For square-lattice SAPs, the counts are now known up to 110 steps \cite{J00},
and the radii of gyration up to 100 steps \cite{J00}.
For triangular-lattice SAPs, the counts $p_N$ were previously
known up to $N=35$ \cite{EG92};
in as-yet-unpublished work, one of us (ANR) has extended them
up to $N=40$ \cite{Rogers_03}, while even more recently another of us (IJ)
has extended the series to $N=60$ steps \cite{J04b}.

As the FLM does not enable us to record the full end-point distribution,
nor higher moments (at least not with the amount of memory available to us),
we also programmed a conventional backtracking algorithm and recorded
the full end-point distribution $c_N({\bf x})$ for $N \le 32$
(square lattice) and for $N \le 22$  (triangular lattice).
As a result, we are able to study arbitrary moments.
We refrain here from inundating the reader
with the complete tables of $c_N({\bf x})$;
they are available on \verb+www.ms.unimelb.edu.au/~iwan+.
However, we do list here most of those series that we subsequently analyse.
For the square lattice, we give in Table~\ref{sq_ri_moments}
the rotationally invariant moments
$\<r^4\>_N$, $\<r^6\>_N$  and $\<r^8\>_N$,
and in Table~\ref{sq_nri_moments}
the corresponding non-rotationally-invariant moments
$\<r^4 \cos 4\theta\>_N$, $\<r^6\cos 4\theta\>_N$ and $\<r^8\cos 4\theta\>_N$.
For the triangular lattice, we give in Table~\ref{tri_ri_moments}
the rotationally invariant moments
$\<r^4\>_N,$  $\<r^6\>_N$ and $\<r^8\>_N$,
and in Table~\ref{tri_nri_moments}
the corresponding non-rotationally-invariant moments
 $\<r^6\cos 6\theta\>_N$  and $\<r^8\cos 6\theta \>_N$.

\subsection{Method of analysis} \label{sec3.2}

In this subsection we explain in detail the method
we used to analyse the data,
the results of which are reported in subsequent subsections.
For the triangular lattice, we expect the series coefficients of any generic
quantity, such as the SAW generating function,
to have an asymptotic expansion of the
form 
\be
\mu^N N^{\gamma-1} \left( a_0 + \sum_{i=1}^k \frac{a_i}{N^{\Delta_i}} \right)
   \;,
\label{m0}
\ee
 where
$\mu$ is the connective constant and $\gamma$ is the critical exponent.  
Likewise, we expect the square-lattice series coefficients to have
an asymptotic expansion of the form 
\be
\mu^N N^{\gamma-1} \left( a_0 + \sum_{i=1}^k \frac{a_i}{N^{\Delta_i}} \right)
   \,+\,
  (-\mu)^N N^{\alpha - 2}
   \left( b_0 + \sum_{i=1}^m \frac{b_i}{N^{\Delta^{\rm AF}_i}}
   \right)
  \;,
\label{mk}
\ee
where $\alpha$ is the critical exponent occurring in the polygon
generating function.  
Similar expansions hold for metric quantities,
and involve also the critical exponent $\nu$.
Since the exact values $\gamma = 43/32$, $\alpha = 1/2$ and $\nu = 3/4$
are well established, we shall use them throughout this paper.

Given the calculated terms of the series up to some order $N_{\rm max}$,
we proceed as follows:
First we decide how many correction terms $\{a_i\}$ and $\{b_i\}$
we wish to include (i.e., we fix the numbers $k$ and $m$);
then we make some assumption for the values of $\mu$, $\Delta_i$ and
$\Delta^{\rm AF}_i$;
finally, we fit the data to \reff{m0} or \reff{mk} by taking
$(k+m+2)$-tuples of successive values of $N$
and solving for $\{a_i\}$ and $\{b_i\}$.
This can be done by solving a system of linear equations.

By using $(k+m+2)$-tuples at steadily larger values of $N$,
many estimates for the $\{a_i\}$ and $\{b_i\}$ are found.
If the different estimates seem stable as $N$ grows,
we presume that they provide an acceptably accurate estimate
of the actual asymptotic coefficients.

A noteworthy feature of the method is that, if a blatantly-too-low
correction-to-scaling exponent is given as input
(for example, specifying $\Delta_1 = 1/2$ for the two-dimensional SAW),
the sequence of amplitude estimates for the term corresponding
to that exponent will converge rapidly to zero, giving a very strong signal
that the exponent in question is absent.
(Of course, if such a term were to occur with an amplitude
several orders of magnitude smaller than the amplitudes of the other terms,
one could be fooled into thinking such a term is absent.
Our analysis assumes the absence of such pathologies.)

Another point to bear in mind is that even if one knows the
precise asymptotic form, with a limited number of series coefficients
one can fit only to a small number of asymptotic terms
(i.e., $k$ and $m$ cannot be taken too large).
Beyond a certain number of terms in the asymptotic form, the quality of
the fit visibly deteriorates.
The more series coefficients are available,
the more terms can be included in the Ansatz
(provided that sufficient numerical precision is retained
during the analysis).

A universally observed feature of the method is that the
apparent accuracy of the amplitude estimates
decreases rapidly as we move to higher-order terms in the asymptotic
expansions. That is to say, the apparent accuracy of the estimate
of amplitude $a_{i+1}$ is significantly less than that of $a_i.$
Moreover, adding further terms in the assumed asymptotic form
(i.e., increasing $k$ and $m$)
improves convergence of the low-order amplitudes $a_i$
{\em provided}\/ that $k+m$ does not get too large,
but after a certain point actually slows the convergence.
In the case at hand, allowing more than 2--5 terms (these being the 
values of $k+1$ and $m+1$ separately)
in the assumed asymptotic form led to a deteriorating
(i.e., less stable) fit. 

As the series data at very small $N$ are probably
not reflective of asymptotic behaviour,
and we have here the luxury of access to many terms
(i.e., quite large $N_{max}$),
the first $19-k-m$ terms of the series will not be used in any
of our analyses here.

Our analysis thus comprises two phases.
In the first phase, we determine the correct connective constant $\mu$
and the correct exponents $\Delta_i$ and $\Delta^{\rm AF}_i$
for the asymptotic expansion, as just described.
In the second phase, we determine
how many terms in the asymptotic expansion we can reliably use.
We now describe our procedure for the second phase of the analysis.

We begin by fitting for only one correction coefficient, $a_1$.
Then we add further asymptotic terms until the estimates obtained
do not appear to be converging as $N\to\infty$ to a value
that is consistent with the previous estimates given by fits with one fewer
asymptotic term.
We define ``consistent'' by the requirement that
estimates of all included asymptotic coefficients be well-converged
and of the same sign and within a factor $F = 2.4$
of the previous estimates.
More specifically, we invoke this requirement as follows:
setting $k=K$ gives estimates of $a_0, \ldots, a_K$;
repeating the analysis with $k=K+1$ yields estimates of
$a_0, \ldots, a_{K+1}$.
We require that the coefficients $a_0, \ldots, a_K$ from the two fits
agree in sign and in magnitude within a factor of $F$;
otherwise, we reject the fit with $k=K+1$ and stop at $k=K$.
Note that the choice of the value of $F$ is somewhat arbitrary.
Realistically, one can reasonably make any choice in the range
$1.5 \ltapprox F \ltapprox 3$, the lower value being more conservative.
We chose a value in this range that included most
data sets with small values of $k$ and $m$, and excluded those with higher
values.

Please note that the convergence (as $N$ grows) of each fit
is here judged by traditional intuitive (and thus somewhat subjective) methods.
It would be an interesting project to find
a precise definition of ``well-converged'' (or its synonym, ``stable'')
that accords satisfactorily with our intuitive judgments
and gives good results on test series;
this would allow the series analysis to be converted into a
precise algorithm.
But we do not purport to carry out such a project here.

For the triangular lattice,
this procedure is thus relatively simple to implement. 
We compute fits initially with $k = 0$,
incrementing $k$ by 1 until a non-stable or inconsistent estimate
(as defined in the preceding paragraph) is found;
we then revert to the previous group of stable and consistent estimates.
The final entries (i.e., those corresponding to the maximum $N$)
in the largest stable group are taken as our final estimates.

For the square lattice, the procedure is more complicated,
as it is not clear {\em a priori}\/ whether terms 
involving $\Delta_i$ or $\Delta^{\rm AF}_i$
should be added to a given group.
Empirically we have found that
groups containing approximately equal numbers of $\Delta_i$ and
$\Delta^{\rm AF}_i$ terms, or slightly more $\Delta_i$ terms,
are more stable than estimates with significantly different values
of $k$ and $m$.
Hence we begin by exploring groups with equal numbers of
$\Delta_i$ and $\Delta^{\rm AF}_i$ terms, that is with $k=m$,
adding one coefficient to each group at every stage.
Next we try groups with one more $\Delta_i$ term than $\Delta^{\rm AF}_i$
terms, so that $k=m+1$;
and finally we try groups with two more $\Delta_i$ terms, so that $k=m+2$.
Again, the largest group that provides
stable and consistent estimates is selected.
As always, the given estimates are taken from the fits to the
largest available value of $N$, which is $N_{max}$,
since these should best reflect the asymptotic regime.

The estimated error is calculated as the change between the estimate
given by the longest series and the series ten terms shorter,
multiplied by a factor reflective of the expected rate of convergence
of the estimates.
This latter factor is determined by assuming that the error
in the estimates is principally given by the first omitted
$\Delta_i$ or $\Delta^{\rm AF}_i$ term.
The difference in the exponents between the term in question
and the first omitted term is then used to
predict the value of the estimate on a fit to an infinite series.

To illustrate these procedures,
we show below the output from fitting the triangular-lattice series
for $c_N \<R_g^2\>_N$
with increasing numbers of correction-to-scaling terms.
We make the Ansatz
\be
(N+1)^2 c_N\<R_g^2\>_N/6  \;\sim\;
   \mu^N N^{123/32}[a_0 + a_1/N + a_2/N^{3/2} + a_3/N^2 + a_4/N^{5/2} + \cdots]
\ee
and obtain fits as follows:

\bigskip

{\small \tt 
\renewcommand{\arraystretch}{0.9}
\begin{tabular}{lll}
 $N$ &	$a_0$	&	$a_1$ \\
 21 & 0.01929438 & 0.08276963 \\
 22 & 0.01932692 & 0.08208623\\
 23 & 0.01935556 & 0.08145624\\
 24 & 0.01938091 & 0.08087327\\
 25 & 0.01940346 & 0.08033194\\
 26 & 0.01942363 & 0.07982768\\
 27 & 0.01944175 & 0.07935660\\
 28 & 0.01945809 & 0.07891535\\
 29 & 0.01947289 & 0.07850104\\
 30 & 0.01948633 & 0.07811114\\
 31 & 0.01949859 & 0.07774345\\
 32 & 0.01950980 & 0.07739603\\
 33 & 0.01952007 & 0.07706717\\
 34 & 0.01952952 & 0.07675536\\
 35 & 0.01953823 & 0.07645924\\
 36 & 0.01954628 & 0.07617761\\
 37 & 0.01955373 & 0.07590938\\
 38 & 0.01956064 & 0.07565359\\
 39 & 0.01956707 & 0.07540935\\
 40 & 0.01957306 & 0.07517587
\end{tabular}
}

\vspace{5mm}

{\small \tt 
\renewcommand{\arraystretch}{0.9}
\begin{tabular}{llll}
 $N$&	$a_0$	  &    $a_1$    &   $a_2$ \\
 22 &  0.01976614 &  0.05375917 &  0.08754307\\
 23 &  0.01976114 &  0.05408172 &  0.08654625\\
 24 &  0.01975680 &  0.05437453 &  0.08562049\\
 25 &  0.01975301 &  0.05464205 &  0.08475611\\
 26 &  0.01974966 &  0.05488785 &  0.08394517\\
 27 &  0.01974669 &  0.05511490 &  0.08318096\\
 28 &  0.01974404 &  0.05532562 &  0.08245789\\
 29 &  0.01974166 &  0.05552207 &  0.08177122\\
 30 &  0.01973951 &  0.05570593 &  0.08111693\\
 31 &  0.01973756 &  0.05587867 &  0.08049156\\
 32 &  0.01973578 &  0.05604149 &  0.07989215\\
 33 &  0.01973415 &  0.05619545 &  0.07931614\\
 34 &  0.01973265 &  0.05634145 &  0.07876131\\
 35 &  0.01973127 &  0.05648027 &  0.07822571\\
 36 &  0.01972999 &  0.05661259 &  0.07770765\\
 37 &  0.01972881 &  0.05673898 &  0.07720563\\
 38 &  0.01972770 &  0.05685998 &  0.07671833\\
 39 &  0.01972667 &  0.05697603 &  0.07624459\\
 40 &  0.01972571 &  0.05708754 &  0.07578337
\end{tabular}
}

\vspace{5mm}

{\small \tt 
\renewcommand{\arraystretch}{0.9}
\begin{tabular}{lllll}
 $N$&	$a_0$	  &    $a_1$    &   $a_2$     & $a_3$ \\
 23 &  0.01970992 &  0.06084150 &  0.04428770 &  0.07428718\\
 24 &  0.01971020 &  0.06080458 &  0.04451851 &  0.07388144\\
 25 &  0.01971034 &  0.06078451 &  0.04464682 &  0.07365079\\
 26 &  0.01971039 &  0.06077784 &  0.04469036 &  0.07357085\\
 27 &  0.01971036 &  0.06078284 &  0.04465705 &  0.07363327\\
 28 &  0.01971026 &  0.06079722 &  0.04455931 &  0.07382009\\
 29 &  0.01971013 &  0.06081948 &  0.04440510 &  0.07412046\\
 30 &  0.01970995 &  0.06084830 &  0.04420187 &  0.07452360\\
 31 &  0.01970976 &  0.06088254 &  0.04395605 &  0.07501986\\
 32 &  0.01970954 &  0.06092130 &  0.04367303 &  0.07560099\\
 33 &  0.01970931 &  0.06096380 &  0.04335757 &  0.07625946\\
 34 &  0.01970908 &  0.06100940 &  0.04301375 &  0.07698862\\
 35 &  0.01970883 &  0.06105753 &  0.04264515 &  0.07778247\\
 36 &  0.01970859 &  0.06110773 &  0.04225488 &  0.07863562\\
 37 &  0.01970834 &  0.06115962 &  0.04184568 &  0.07954324\\
 38 &  0.01970809 &  0.06121284 &  0.04141995 &  0.08050094\\
 39 &  0.01970785 &  0.06126712 &  0.04097980 &  0.08150474\\
 40 &  0.01970761 &  0.06132220 &  0.04052709 &  0.08255105
\end{tabular}
}

\vspace{5mm}

{\small \tt 
\renewcommand{\arraystretch}{0.9}
\begin{tabular}{llllll}
 $N$&	$a_0$	  &    $a_1$    &   $a_2$     &   $a_3$	    &  $a_4$\\
 24 &  0.01971252 &  0.06028314 &  0.04946255 &  0.05630611 &  0.02220967\\
 25 &  0.01971161 &  0.06048759 &  0.04752405 &  0.06319720 &  0.01350151\\
 26 &  0.01971081 &  0.06067477 &  0.04571025 &  0.06978710 &  0.00499021\\
 27 &  0.01971004 &  0.06086343 &  0.04384348 &  0.07671271 & -0.00414364\\
 28 &  0.01970935 &  0.06103877 &  0.04207338 &  0.08341263 & -0.01315896\\
 29 &  0.01970871 &  0.06120830 &  0.04032868 &  0.09014499 & -0.02239437\\
 30 &  0.01970812 &  0.06137074 &  0.03862565 &  0.09683963 & -0.03175015\\
 31 &  0.01970757 &  0.06152632 &  0.03696505 &  0.10348538 & -0.04120539\\
 32 &  0.01970707 &  0.06167592 &  0.03534052 &  0.11010004 & -0.05078048\\
 33 &  0.01970660 &  0.06181962 &  0.03375373 &  0.11666979 & -0.06045075\\
 34 &  0.01970616 &  0.06195787 &  0.03220232 &  0.12319772 & -0.07021604\\
 35 &  0.01970575 &  0.06209097 &  0.03068515 &  0.12968223 & -0.08006943\\
 36 &  0.01970536 &  0.06221921 &  0.02920108 &  0.13612229 & -0.09000497\\
 37 &  0.01970501 &  0.06234286 &  0.02774880 &  0.14251786 & -0.10001832\\
 38 &  0.01970467 &  0.06246219 &  0.02632719 &  0.14886860 & -0.11010477\\
 39 &  0.01970435 &  0.06257743 &  0.02493510 &  0.15517452 & -0.12026033\\
 40 &  0.01970406 &  0.06268879 &  0.02357145 &  0.16143576 & -0.13048133
\end{tabular}
}

\vspace{5mm}

We observe in these fits the following behaviour:
As $N$ increases, the estimates of the amplitudes ${a_i}$
appear to be converging in each set, until we reach
the set with five asymptotic coefficients ($a_0,\ldots,a_4$).
In this latter fit, we see that the estimate of $a_4$ appears to be diverging.
Further, the estimates of $a_1,$ $a_2$ and $a_3$ have deteriorating apparent
convergence as we go from a four-term to a five-term fit.
By contrast, going from a two-term to a three-term fit,
and from a three-term to a four-term fit,
improved the apparent convergence of the amplitude sequences.
Thus we reject the five-term fit, and base our estimates on the
four-term fit.

Finally, the estimated error from a series of length $N$ is taken to be the
appropriately scaled difference between the values obtained from this series
and those obtained from the series of length $N - 10$.
This difference is scaled by a factor dependent on
the difference between the exponent in question
and the first omitted exponent.
The scaling factor follows from our assumption that
the error is given principally by the first neglected term, 
$c/N^{\Delta_{k+1}}$ (or similarly with $\Delta^{\rm AF}_{m+1}$). 
Hence, if the actual value of the coefficient in question is $a_i$ and the
two estimates are $a_i^{(N)}$ and $a_i^{(N-10)}$, we expect that
\begin{subeqnarray}
\frac{a_i^{(N)}}{N^{\Delta_{i}}}& =&
    \frac{a_i}{N^{\Delta_{i}}} + \frac{c}{N^{\Delta_{k+1}}} \\[2mm]
\frac{a_i^{(N-10)}}{(N-10)^{\Delta_{i}}} &= &
    \frac{a_i}{(N-10)^{\Delta_{i}}} +
    \frac{c}{(N-10)^{\Delta_{k+1}}}
\end{subeqnarray}
Simple algebra then yields
\be
a_i^{(N)}-a_i  \;=\;
  - \frac{a_i^{(N)}-a_i^{(N-10)}}
         {\bigl(\frac{N}{N-10}\bigr)^{\Delta_{k+1}-\Delta_{i}} - 1}
   \;.
\ee
Therefore, $a_i$ is estimated by $a_i^{(N)}$ with error quoted as 
\be
   2 |a_i^{(N)}-a_i^{(N-10)}| /
   [(\frac{N}{N-10})^{\Delta_{k+1}-\Delta_{i}} - 1]  \;.
\ee
The factor of 2 is included to make our errors more conservative.
More adventurous readers may choose to reduce this factor.

In the above example, the first omitted term is O$(N^{-5/2})$.
The difference in the estimate of $a_0$ from the $N=40$ series
with four asymptotic coefficients (0.019707611)
and that from a $N=30$ series (0.019709959) is 2.348 $\times 10^{-6}$.
Thus the error is quoted as
$2 \times 2.348 \times 10^{-6}/((40/30)^{2.5} - 1)
=4.5 \times 10^{-6}.$
Our amplitude estimate is then $a_0 = 0.019708 \pm 0.000005.$
Similarly, $a_1 = 0.0613 \pm 0.0018,$ where the error is given by
$2 \times 0.0004739 /((40/30)^{1.5} - 1).$
Likewise, $a_2 = 0.04 \pm 0.02,$ and $a_3 = 0.08 \pm 0.1.$

\subsection{SAW counts}\label{sec3.3}

In this subsection we discuss the analysis of
the newly extended series for SAW counts on the
square and triangular lattices.
Here we give only a brief analysis,
as fuller details will be published elsewhere \cite{J04a},
along with a discussion of the series derivation.

Let us begin with the triangular lattice.
We first analysed the extended SAP series
using biased differential approximants using the known exponent $\alpha = 1/2$
(see \cite{JG99} for the method).
We obtained 
\be 
   x_c \;=\; 1/\mu \;=\;  0.240\,917\,574 \pm 0.000\,000\,004  \; ,
 \label{eq.tri.xc}
\ee
which we will use in subsequent analyses.

We also performed a similar analysis using the extended SAW series,
biasing the estimate with the known exponent $\gamma = 43/32.$ 
We obtained the estimate
   $x_c \;=\; 1/\mu \;=\;  0.240\,917\,579 \pm 0.000\,000\,008,  \;$
in agreement with the SAP result but less precise.

Using the estimate \reff{eq.tri.xc} of $x_c$,
we proceeded as described in Section~\ref{sec3.2}
to fit the series coefficients to various asymptotic forms.
For triangular-lattice SAWs,
we expect, based on earlier investigations of the
corresponding square-lattice series~\cite{CG96}, that
\begin{equation}\label{asy1}
c_N  \;\sim\;
    \mu^N N^{11/32}[a_0 + a_1/N + a_2/N^{3/2} + a_3/N^2 + a_4/N^{5/2} +
a_5/N^3 +   \cdots] \;.
\end{equation}
However, as discussed in Section~\ref{sec2.2},
renormalization-group theory predicts an additional
``energy-like'' term arising from the mixing between
nonlinear scaling fields.
For the two-dimensional SAW ($\alpha = 1/2$),
incorporating this term requires that \reff{asy1} be modified to read
\begin{equation}\label{asy2}
c_N  \;\sim\;
    \mu^N N^{11/32}[a_0 + a_1/N + a_2/N^{3/2} + \widetilde{a}_0/N^{59/32}
       + a_3/N^2 + a_4/N^{5/2} + \cdots]
\end{equation}
[cf.\ \reff{eq.CN.modified}].

In our analysis, we tried both the asymptotic forms
\reff{asy1} and \reff{asy2}.
Since the exponents associated with amplitudes $\widetilde{a}_0$ and $a_3$
are numerically close (1.84375 and 2, respectively),
we expect that it will be very difficult
to distinguish numerically between the Ans\"atze \reff{asy1} and \reff{asy2}.
This is indeed the case.
Under the assumption~\reff{asy1}, we found that the sequences corresponding to
the amplitudes are well converged up to $k=4$ 
for triangular-lattice SAW and
we estimate $a_0 = 1.183966(1),$ $a_1 = 0.5960(4),$
$a_2 = -0.274(6),$ $a_3 = -0.14(4),$ and $a_4 = 0.09(10).$
Our errors are calculated as described in Section~\ref{sec3.2}
and are given, in parentheses, as the 
uncertainty in the last quoted digit(s).
Under the alternative Ansatz \reff{asy2}, we find that
the fit is neither better nor worse.
We observed that the sequences of estimates
of the corresponding amplitudes $\widetilde{a}_0$ and $a_3$
appear to be correlated:
they are monotonically increasing in magnitude but
are of opposite sign, the sum $\widetilde{a}_0 + a_3$ being almost constant.

To investigate this point further, we constructed a test series, with known
asymptotic behaviour, similar to that in \reff{asy2}, namely
\begin{equation}\label{test}
d_N  \,=\, N^{11/32} [1 + 1/N + 0.7/N^{3/2} + 1.25/N^{59/32}
                      + 3/N^2 - 4/N^{5/2} + 5/N^3 -6/N^{7/2}] + (0.5)^N .
\end{equation}
The last term is included to incorporate the fact that there are other
singularities in the
complex plane, beyond $x_c$, which will make an exponentially decaying
contribution to the asymptotics.
We generated the first 1000 terms of this sequence
and analysed them as above,
including either a term $N^{-59/32}$ or a term $N^{-2}$ or both.
The analyses using either one of these two terms behaved similarly.
The analysis using both terms gave {\em inferior}\/ estimates
of the first three amplitudes, and the {\em wrong sign}\/
for the amplitude of the $N^{-2}$ term, when $N \ltapprox 240$.
Only beyond this point does the analysis using both terms
give superior estimates of the first three amplitudes,
along with the right sign for the $N^{-2}$ term
(the two issues clearly go together).
We conclude that using series of the length available to us
($N \ltapprox 40$), it is unfeasible to determine
whether a term $N^{-59/32}$ is present or absent.

In conclusion, our analysis is unable to
resolve the question of whether the ``energy-like'' term $N^{-59/32}$
is present or not.  Therefore, for the
subsequent analysis of the metric quantities,
reported in the next subsection, we have assumed for simplicity
the absence of this term,
and just assumed the asymptotic form~\reff{eq_wegner} with one
correction-to-scaling exponent $\Delta_1 = 3/2.$

On the square lattice, the situation is complicated by the presence of
an ``antiferromagnetic'' singularity at $x = -1/\mu$.
{}From \reff{corr_AF} ff.\ we recall that
the asymptotic form of the coefficients given in~\reff{asy1} and~\reff{asy2}
is modified by the additional term
\begin{equation}\label{afterm}
(-1)^N \mu^N N^{-3/2}[d_0 + d_1/N + d_2/N^2 + d_3/N^3 + \cdots].
\end{equation}
Analysing the square-lattice SAW data using as our estimate of
$x_c $ the positive real root of 
\be\label{sqxc}
581x^4 + 7x^2 - 13 = 0,
\ee
which is a useful mnemonic for the current best estimate
$x_c^2 = 0.143\,680\,629\,27(1)$ \cite{JG99},
we obtain a very convincing fit with $k = 3$ and $m = 2$,
enabling the following amplitude estimates to be made:
$a_0 = 1.1770425(7),$ $a_1 = 0.5501(2),$ $a_2 = -0.1402(3)$ and $a_3 = -0.12(2)$ for the
``ferromagnetic'' amplitudes,
and $d_0 = -0.189848(3),$ $d_1 = 0.17473(9)$ and $d_2 = -1.51(1)$
for the ``antiferromagnetic'' amplitudes.
This is in close agreement with the earlier estimates \reff{eq1.1}
based on slightly shorter series;
here we have obtained a slight improvement in the precision of
the estimates for the leading amplitudes $a_0,$ $d_0$ and $d_1$.

If instead we assume the asymptotic form \reff{asy2} for the
``ferromagnetic'' term, we find that estimates of $a_3$ are
small (less than 0.03 in magnitude) and tending toward zero.
Estimates of $a_4$ are tending toward the estimate for this
term obtained in the absence of the additional term with
amplitude $a_3.$
Once again, we are unable to distinguish \reff{asy1} from \reff{asy2}.

\subsection{Metric quantities $\<R_e^2\>$, $\<R_g^2\>$ and $\<R_m^2\>$}
  \label{sec3.4}

In this subsection we shall analyse
the metric quantities $\<R_e^2\>_N$, $\<R_g^2\>_N$ and $\<R_m^2\>_N$,
any one of which we shall generically denote by $\<R^2\>_N$.
As discussed in Section~\ref{sec2.2}
[cf.\ \reff{eq_wegner}/\reff{corr_AF}],
their asymptotic behaviour is expected in the first instance to be
\be
\<R^2\>_N  \;\sim\;
     N^{2\nu}[a_0 + a_1/N + a_2/N^{\Delta_1} + a_3/N^2 + \cdots]
  \label{6tri}
\ee
for the triangular lattice, and
\begin{eqnarray}
\<R^2\>_N  & \sim &
     N^{2\nu}[a_0 + a_1/N + a_2/N^{\Delta_1} + a_3/N^2 + \cdots]
        \nonumber \\
     & &  +\; (-1)^N N^{q}[b_0 + b_1/N^{\Delta_1^{\rm AF}} + b_2/N + \cdots]
  \label{6}
\end{eqnarray}
for the square lattice.
As mentioned earlier, there is overwhelming numerical evidence
\cite{Li-Madras-Sokal,CG96}
that the leading exponent $2\nu$ equals 3/2 exactly,
as predicted by Coulomb-gas arguments \cite{Nienhuis_82,Nienhuis_84};
we shall henceforth take this fact for granted.
We also expect $\Delta_1 = 3/2$,
as predicted by Nienhuis~\cite{Nienhuis_82}
and confirmed numerically for the SAW counts.
Furthermore, for the square lattice we predict $q = -11/32$
[cf.\ \reff{q2a}].
Finally, in Section~\ref{sec2.2} we pointed out that
renormalization-group theory
predicts an additional ``energy-like'' term in the susceptibility
[cf.\ \reff{asy2}],
though alas we were unable to distinguish it numerically
(see Section~\ref{sec3.3});
it is reasonable to guess that there may be a corresponding term
also in the series for the unnormalized second moments
$c_N \<R^2\>_N$.
Whether or not the latter term is present,
the existence of an ``energy-like'' term in $c_N$
will induce in $\<R^2\>_N$ additional correction-to-scaling terms
$N^{-59/32}$, $N^{-91/32}$, \ldots\ 
beyond those included in \reff{6tri}/\reff{6}.

In addition to the normalized metric quantities $\<R^2\>_N$,
we also studied the corresponding unnormalized quantities $c_N \<R^2\>_N$,
whose expected asymptotic form is
\be
 c_N  \<R^2\>_N  \;\sim\;
   \mu^N  N^{2\nu+\gamma-1}[b_0 + b_1/N + b_2/N^{\Delta_1} + b_3/N^2 + \cdots]
   \;,
  \label{7tri}
\ee
with appropriate additional antiferromagnetic terms \reff{q2} 
when analysing the square-lattice data.
The latter quantities have the disadvantage that the analysis
depends sensitively on an input estimate of $\mu$;
but, for loose-packed lattices and for
$\<R_e^2\>_N$ only, 
they have the advantage 
that the effect of the antiferromagnetic singularity is weaker.
To see this, compare \reff{q1}--\reff{q2a}:
the antiferromagnetic contribution in $c_N \<R_e^2\>_N$
is relatively weaker than that in $c_N$;
but the antiferromagnetic contribution in $\<R_e^2\>_N$
is dominated by that in $c_N$.
Therefore, the antiferromagnetic contribution is
relatively weaker in $c_N \<R_e^2\>_N$ than in $\<R_e^2\>_N$.

Our method of analysis is based on directly fitting $\<R^2\>_N$
and $c_N \<R^2\>_N$
to the assumed asymptotic form \reff{6tri}/\reff{6}/\reff{7tri}, 
as described in Section \ref{sec3.2}.
The values of the exponents $\nu$, $q$, $\Delta_1$
and $\Delta_1^{\rm AF}$ are assumed, and the appropriate system of
linear equations is solved to give estimates of the amplitudes
$\{a_i\}$ and $\{b_i\}$.
In applying the method to metric quantities (see, for example, the table in
Section \ref{sec3.2}), the fit to the
leading amplitude is rather stable, that to the first analytic
correction term is moderately stable, while the fit to the amplitude of the
assumed correction-to-scaling term $N^{-3/2}$
converges less impressively for both the normalized and unnormalized
metric quantities.
As already noted, adding further terms in the assumed asymptotic form
beyond the first initially improved convergence, but this improvement is not
 sustained. That is to say, allowing more than between two and five terms in 
the assumed asymptotic form led to an apparently deteriorating fit.

\subsubsection{Triangular lattice}
With our 40-term triangular-lattice series we found
that we could fit to
$a_0$, $a_1$, $a_2$ and sometimes $a_3$.
For the normalized and unnormalized metric quantities,
the estimate of $a_0$ could usually be made to four-digit precision,
while the estimate of $a_1$ could be made only to one or two significant
digits, and the estimate of $a_2$
is accurate only to at best one significant digit.
For $a_3$ the error is comparable to or greater than the estimate.

We have applied this analysis method
to the triangular-lattice data,
using a 40-term series for all metric quantities.
Because the triangular lattice is close-packed,
there is only one singularity on the circle of convergence,
which makes the analysis simpler than for the square lattice
[compare \reff{6tri} to \reff{6}].

The tables of estimates for the metric quantities obtained according
to the procedure described in Section \ref{sec3.2}
are shown in Table~\ref{tableR2t}.
The results are:
\begin{eqnarray}
\<R_e^2\>_N   & \sim &
      N^{3/2}[0.71174(32) + 0.95(12)/N - 2.6(1.6)/N^{3/2} +  3(7)/N^2 + {\rm O}(1/N^{5/2})]
  \nonumber \\ \label{17b}  \\[2mm]
\<R_g^2\>_N   & \sim &
      N^{3/2}[0.09989(4) + 0.056(16)/N + 0.3(2)/N^{3/2} - 0.2(1.0)/N^2 +  {\rm O}(1/N^{5/2})]
  \nonumber \\ \label{18b}  \\[2mm]
\<R_m^2\>_N   & \sim &
      N^{3/2}[0.3133(4) + 0.24(12)/N - 0.2(1.0)/N^{3/2} + {\rm O}(1/N^2) ]
  \label{19b}.
\end{eqnarray}

These were obtained with $k=3,$ $k=3,$ and $k=2$ respectively.
Unfortunately the uncertainties in the coefficients of the O$(1/N^{3/2})$
are so great as to be comparable to (or, in the case of $\<R_m^2\>$, larger
than) the coefficient itself. 
Further, the analysis of the $\<R_g^2\>_N$ series violates the convergence
criterion we have set, in that the coefficient of $a_1$ differs by nearly
a factor of 3 in going from a two-term fit ($k = 1$) to a three-term fit
($k = 2$). We have nevertheless presented results for $k =3.$
Our justification for this is twofold. Firstly,
the estimate of $a_1$ stabilises if we then go to a four-term fit.
Secondly, as we have already seen, the data for the essentially
equivalent series $(N+1)^2 c_N\<R_g^2\>_N/6$
supports a four-term fit. For the reader unconvinced by these arguments,
the corresponding analysis with $k = 0$ (a one-term fit, as would be justified
by strict adherence to the convergence criteria we have set), gives 
$a_0 = 0.106 \pm 0.012.$

We can do somewhat better from
a similar analysis of the unnormalized metric quantities,
using the estimate $\mu = 4.15079723$ from \reff{eq.tri.xc},
which gave
\begin{eqnarray}
c_N\<R_e^2\>_N/6  & \sim &
      \mu^N N^{59/32}[0.14045(6) + 0.256(26)/N - 0.53(32)/N^{3/2}  \nonumber \\
&   & + 0.6(1.6)/N^2 + {\rm O}(1/N^{5/2}) ]
  \label{17a}  \\[2mm]
(N+1)^2 c_N\<R_g^2\>_N/6  & \sim &
      \mu^N N^{123/32}[0.019708(5) + 0.0613(18)/N + 0.04(2)/N^{3/2} \nonumber \\
&   & +  0.08(10)/N^2 + {\rm O}(1/N^{5/2}) ]
      \label{18a}  \\
(N+1) c_N\<R_m^2\>_N/6   & \sim &
      \mu^N N^{91/32}[0.06183(12) + 0.136(28)/N - 0.02(24)/N^{3/2} + {\rm O}(1/N^{2}) ]
\nonumber \\
  \label{19a}.
\end{eqnarray}

{}These estimates were also obtained with $k=3,$ $k=3$ and $k=2$ respectively.
{}From these we can estimate the amplitudes of the metric quantities
by dividing through by the asymptotic form \reff{asy1} for $c_N$, and
accounting for
the factor of 6 and appropriate factors of $(N+1).$ In this way we obtain
a second set of amplitude estimates,
\begin{eqnarray}
\<R_e^2\>_N   & \sim &
      N^{3/2}[0.71176(15) + 0.94(7)/N - 2.5(8)/N^{3/2} + 3(3)/N^2 + {\rm O}(1/N^{5/2}) ]
  \nonumber \\ \label{17}  \\[2mm]
\<R_g^2\>_N   & \sim &
      N^{3/2}[0.09987(2) + 0.061(5)/N + 0.23(5)/N^{3/2} - 0.5(5)/N^2 + {\rm O}(1/N^{5/2}) ]
 \nonumber \\ \label{18}  \\[2mm]
\<R_m^2\>_N   & \sim &
      N^{3/2}[0.3133(3) + 0.22(7)/N - 0.03(60)/N^{3/2} + {\rm O}(1/N^2) ]
  \label{19}.
\end{eqnarray}
These differ from the directly measured amplitudes only within the quoted errors
for each amplitude, consistent with our claimed errors.

One immediate observation is that for $\<R_e^2\>_N$ and $ \<R_m^2\>_N$
the correction-to-scaling amplitudes corresponding
to the $1/N$ and $1/N^{3/2}$ terms are of {\em opposite}\/ sign,
while for $ \<R_g^2\>_N$ they are of the the same sign
(though the errors associated with the estimates
of the amplitude of the $1/N^{3/2}$ term are rather large).
Note too that for both $\<R_e^2\>_N$ and $\<R_g^2\>_N$
the amplitude of the $1/N^{3/2}$ term is larger (in magnitude)
than the amplitudes of both the leading term and first analytic correction;
for $\<R_m^2\>$, by contrast, the error in the $1/N^{3/2}$ term
is too great to comment on the relative size of this term.

As a consequence, the ``effective'' exponent $\Delta_{\rm eff}$
based on fitting to a given range of $N$ behaves differently
as a function of $N$ for the different observables.
For $\<R_g^2\>_N$, $\Delta_{\rm eff}$ lies between 1 and 3/2 for all $N$,
and decreases monotonically to 1 as $N \to\infty$.
For $\<R_e^2\>_N$ and $ \<R_m^2\>_N$, by contrast,
$\Delta_{\rm eff}$ starts {\em above}\/ 3/2 for very small $N$,
then {\em increases}\/ monotonically,
reaching $+\infty$ at some finite value of $N$ (here $\approx 15$);
then it jumps to $-\infty$,
after which it continues to increase monotonically,
tending asymptotically to 1 as $N \to\infty$.
These observations are in accordance with previous studies.
Most studies of $\<R_e^2\>_N$ resulted in estimates for $\Delta$ of
$\approx 0.65$
\cite{Djordjevic_83,Privman_84,Ishinabe_88, Ishinabe_89,Lam_90},
while most studies of $\<R_g^2\>_N$ resulted in estimates for $\Delta$
in the range 1.05--1.2 \cite{Ishinabe_88, Ishinabe_89,Havlin_83, Lam_90}.
(There have been few previous studies of $\<R_m^2\>_N$.)
This is clearly a source---indeed, probably the major source---of
the long-standing difficulty in the analysis of these quantities
for the correction-to-scaling term.

The amplitude ratios $A$ and $B$
[defined as the $N\to\infty$ limit of \reff{ratios}]
follow immediately from~\reff{17}--\reff{19}
as $A = 0.14031$ and $B = 0.4398$,
and from~\reff{17b}--\reff{19b}
as $A = 0.14033$ and $B = 0.4402$.

We also estimated the ratios $A$ and $B$ by direct extrapolation
of the appropriate coefficient quotients, using the following method
\cite{OPBG}:
Given a sequence $\{a_n\}$ defined for $n \ge 1,$
which is known or assumed to converge to a limit $a_{\infty}$
with corrections of the form
$a_n \sim a_{\infty}(1 + b/n + \ldots)$,
we first construct a new sequence $\{h_n\}$
defined by $h_n = \prod_{m=1}^n a_m$.
Then the generating function
$\sum h_n x^n \sim (1 - a_{\infty} x)^{-(1+b)}$.
We obtain estimates of the required limit $a_{\infty}$
and parameter $b$ by analysing this generating
function by the standard method of differential approximants.
(The value of the parameter $b$ can also be obtained numerically
 from the amplitude estimates given in \reff{17}--\reff{19} above.)
In this way, we obtain the estimates
\begin{eqnarray}
    A  & = &  0.140296(6)  \label{tA} \\
    B  & = &  0.439649(9)\; .  \label{tB}
\end{eqnarray}

\subsubsection{Square lattice}  \label{sec3.5.2}

Let us now consider the square-lattice data.
We first analysed the three metric quantities
$\<R_e^2\>_N$, $\<R_g^2\>_N$ and $\<R_m^2\>_N$
by a method similar to that leading to \reff{17b}--\reff{19b},
but including the contribution of the antiferromagnetic singularity.
We imposed the exponent values
$\nu = 3/4$, $\Delta_1 = 3/2$, $q = -11/32$ and $\Delta_1^{\rm AF} = 1$;
the justification for these choices has already been given above.
The sequences of amplitude estimates
are shown in Tables \ref{tableR2es}, \ref{tableR2gs} and \ref{tableR2ms}.
In this way, we obtain the following results:
As discussed in Section (3.2), some experimentation was needed to determine
the maximum values of the parameters $m$ and $k$ in \reff{mk}.
The results given below are given by
$m=4,$ $k=2$ for $\<R_e^2\>_N,$
by $m=2,$ $k=1$ for $\<R_g^2\>_N,$
and by $m=3,$ $k=1$ for $\<R_m^2\>_N.$
We find
\begin{eqnarray}
\<R_e^2\>_N  &\sim & N^{3/2}[0.77124(5) + 1.159(38)/N - 3.13(74)/N^{3/2} + 6(6)/N^2 - 6(24)/N^{5/2} 
\nonumber \\
&+& 0.4(4.0)/N^3 + {\rm O}(1/N^{7/2})] + (-1)^N N^{-11/32}[0.12451(17) - 0.027(24)/N + {\rm O}(1/N^2)]
      \nonumber \\ \label{sqfita}   \\[3mm]
\<R_g^2\>_N  & \sim & N^{3/2}[0.108230(1) + 0.1019(1)/N + 0.1082(4)/N^{3/2}
-{\rm O}(1/N^2)]
      \nonumber \\
 &  & +\; (-1)^N N^{-11/32}[0.008364(19) + 0.0031(21)/N +{\rm O}(1/N^2)]
      \label{sqfitb}   \\[3mm]
\<R_m^2\>_N  &\sim & N^{3/2}[0.33913(8) + 0.426(17)/N - 1.1(1.1)/N^{3/2} + 2(4)/N^2 + {\rm O}(1/N^{5/2})]
      \nonumber \\
 &  & +\; (-1)^N N^{-11/32}[0.03652(11) + 0.015(12)/N + {\rm O}(1/N^2)]\; .
      \label{sqfitc}
\end{eqnarray}
A similar analysis of the unnormalized quantities, using the estimate\\
 $\mu=2.63815853034174086843$
from \reff{sqxc}, was made.
The results below are given by
$m=4,$ $k=2$ for  the first two quantities, and
$m=3$, $k=1$ for the third.
For the first two quantities we have not given our estimate of $a_4$ as 
the associated error is significantly bigger than the estimate.
We find
\begin{eqnarray}
c_N\<R_e^2\>_N/4  &\sim & N^{59/32}[0.226945(14) + 0.4471(11)/N - 0.95(22)/N^{3/2} + 2(2)/N^2 + {\rm O}(1/N^{5/2})]
      \nonumber \\
 &  & +\; (-1)^N N^{-3/2}[0.019098(1) + 0.0415(41)/N - 0.08(45)/N^2 + {\rm O}(1/N^3)]
      \label{sqfitac}   \\[3mm]
c_N\<R_g^2\>_N  & \sim & N^{59/32}[0.127388(31) + 0.181(17)/N + 0.10(26)/N^{3/2} + 0.1(1.0)/N^2 + {\rm O}(1/N^{5/2})]
      \nonumber \\
 &  & +\; (-1)^N [-0.010688(15) + 0.0047(17)/N  - 0.20(5)/N^2 + {\rm O}(1/N^3)]
      \label{sqfitbc}   \\[3mm]
c_N\<R_m^2\>_N  &\sim & N^{59/32}[0.39917(11) + 0.686(44)/N - 1.3(6)/N^{3/2} + 2(2)/N^2 + {\rm O}(1/N^{5/2})]
      \nonumber \\
 &  & +\; (-1)^N [-0.021383(45) + 0.028(5)/N + {\rm O}(1/N^2)]\; .
      \label{sqfitcc}
\end{eqnarray}
{}From these we can estimate the amplitudes of the normalized metric quantities by dividing
through by the asymptotic form \reff{q1} for $c_N$. In this way we obtain a second set of amplitude estimates,
\begin{eqnarray}
\<R_e^2\>_N  &\sim & N^{3/2}[0.77124(5) + 1.1593(38)/N - 3.13(75)/N^{3/2} + 6(8)/N^2 + {\rm O}(1/N^{5/2})]
      \nonumber \\
 &  & +\; (-1)^N N^{-11/32}[0.12439(4) - 0.0144(9)/N + {\rm O}(1/N^2)]
      \label{sqfitar}   \\[3mm]
\<R_g^2\>_N  & \sim & N^{3/2}[0.108227(50) + 0.103(17)/N + 0.098(160)/N^{3/2}
-{\rm O}(1/N^2)]
      \nonumber \\
 &  & +\; (-1)^N N^{-11/32}[0.008376(20) + 0.0006(30)/N +{\rm O}(1/N^2)]
      \label{sqfitbr}   \\[3mm]
\<R_m^2\>_N  &\sim & N^{3/2}[0.33913(8) + 0.424(19)/N - 1(1)/N^{3/2} + {\rm O}(1/N^2)]
      \nonumber \\
 &  & +\; (-1)^N N^{-11/32}[0.03654(7) + 0.027(8)/N + {\rm O}(1/N^2)]\; .
      \label{sqfitcr}
\end{eqnarray}
These differ from the directly analysed amplitudes only in the last quoted digits
for all but the least significant amplitudes, and are consistent with our quoted
errors in all cases.
In the notation of \reff{rel-a0d0},
from \reff{sqfitar} we have $a''_0 / d''_0 = 6.200(3)$,
while from the amplitudes quoted below \reff{sqxc}
we have $a_0 / d_0 = 6.1999(1)$, in complete agreement.

The amplitude ratios $A$ and $B$
follow immediately from~\reff{sqfita}--\reff{sqfitc} as
$A = 0.14033$ and $B = 0.43971$.
{}From~\reff{sqfitar}--\reff{sqfitcr} we obtain the almost identical values,
$A = 0.14033$ and $B = 0.43972$.

We also analysed these amplitude ratios directly, using the
same method as discussed above for the analysis of the
triangular-lattice data. We obtained
\begin{eqnarray}
    A  & = &  0.140299(6)   \label{sqA} \\
    B  & = &  0.439647(6)\; . \label{sqB}
\end{eqnarray}
Comparison with the corresponding estimates \reff{tA}--\reff{tB}
for the triangular lattice
is entirely consistent with the belief that these ratios
are lattice-independent \cite{CG93}.

These amplitude ratios are also consistent with the CSCPS relation \reff{CSC}:
using our best estimates \reff{sqA}--\reff{sqB},
we find $F \equiv \lim_{N \rightarrow \infty} F_N = -0.000024 \pm 0.000025$
for the square lattice;
and using \reff{tA}--\reff{tB},
we find $F \equiv \lim_{N \rightarrow \infty} F_N = -0.000036 \pm 0.000036$
for the triangular lattice.
A direct analysis of the sequence $\{F_N\}$ was also undertaken,
but that sequence was found difficult to extrapolate;
and our estimate of the limit,
while entirely consistent with zero,
was a factor of 10 less precise than the one just quoted.

If it is in fact true (as certainly seems to be the case)
that $F_N \to 0$ as $N \to\infty$,
then it is of interest to investigate the {\em rate}\/ at which $F_N \to 0$.
If $F_N \propto N^{-\delta}$,
then $f_N \equiv  F_N \< R^2_e\>_N  \propto N^{3/2-\delta}$.
For both square and triangular lattices, we find that $\delta = 3/2$,
i.e.\ that $f_N$ approaches a nonzero constant
$f \equiv \lim_{N \rightarrow \infty} f_N$ as $N \to\infty$.
This behaviour was initially surprising to us, because it implies that
there is no analytic correction-to-scaling term $1/N$ in $f_N$,
even though such terms are manifestly present in each of the three
individual metric quantities $\< R^2 \>$.
Moreover, this remarkable result appears to hold
for both lattices.\footnote{
   Indeed, as shown in \cite{OPBG} it appears to hold
   even in the case of {\em interacting}\/ SAWs
   within the good-solvent regime (i.e., above the theta temperature).
   Of course, the limiting constant $f$ depends on the interaction.
   For repulsive nearest-neighbour interactions,
   $f$ increases from 0.78 to an asymptotic value of
   about 1.6 as the repulsion gets very strong.
}
However, we were subsequently able to provide a renormalization-group
argument for this cancellation (see Section~\ref{sec2.2} above).
Our estimates of the amplitude are
$f = 0.78 \pm 0.03$ \cite{BW98} for the square lattice
and $f = 0.96 \pm 0.04$ for the triangular lattice.
These estimates are based on extrapolation of the sequences $f_N$
using a variety of extrapolation algorithms,
including Levin's {\em u} transform, Brezinski's $\theta$ algorithm,
Neville-Aitken extrapolation and Wynn's $\epsilon$ algorithm.
Details of these and other algorithms, as well
as programs for their implementation, can be found in \cite{Guttmann_89}.

As noted in Section~\ref{sec2.2},
the observation that $\delta = 3/2$
implies the constraint \reff{eq_CSCPS_subdominant}
on the subdominant amplitudes arising in \reff{27a}--\reff{27c}.
%
Our series estimates \reff{17}--\reff{19} and \reff{sqfita}--\reff{sqfitc}
are consistent with this prediction,
as are our Monte Carlo estimates \reff{4.3}--\reff{4.8}.
Furthermore, our series and Monte Carlo estimates of $f$
are consistent with the relation \reff{eq_CSCPS_f};
but the associated error bars are very large, so this is not a stringent test.

Finally, we note the fact that $\delta = 3/2$ is another indicator that
the correction-to-scaling exponent is indeed $3/2$.
If it were less than this, then the leading non-analytic
correction-to-scaling term would have to cancel miraculously
(as the $1/N$ term does) in the combination \reff{fN-def} for $f_N$.
This seems {\em a priori}\/ unlikely.

\subsection{Euclidean-invariant moments of the distribution function}
   \label{sec3.5}

We have also analysed the series for rotationally-invariant
and non-rotationally-invariant moments of the endpoint distribution function,
given in Tables \ref{sq_ri_moments}--\ref{tri_nri_moments},
using methods similar to those just described for the analysis of $\< R^2 \>$.
Let us start with the rotationally-invariant moments $\<r^{2k}\>_N$,
for which we expect an asymptotic behaviour of the form
\begin{eqnarray}
   \label{rrt_apriori}
\<r^{2k} \>_N
 & \sim &  N^{2k\nu} [c_{0,k} + c_{1,k}/N + c_{2,k}/N^{\Delta_1}
    + c_{3,k}/N^2 + \ldots ]
\end{eqnarray}
for the triangular lattice, and
\begin{eqnarray}
   \label{rrs_apriori}
\<r^{2k} \>_N
& \sim & N^{2k\nu} [c_{0,k} + c_{1,k}/N + c_{2,k}/N^{\Delta_1}
    + c_{3,k}/N^2 + \ldots] \nonumber \\
&  &  +\; (-1)^N N^{q_k}[d_{0,k} + d_{1,k}/N^{\Delta_1^{\rm AF}}
    + d_{2,k}/N + \ldots]
\end{eqnarray}
for the square lattice.
In Section~\ref{sec2.2} we gave arguments predicting that
$q_k = 2k\nu + \alpha - 1 - \gamma = 3k/2 - 59/32$.
Furthermore, it is reasonable to expect $\Delta_1^{\rm AF} = 1$
as was already observed for the SAW counts \cite{CG96}
and for the metric quantities $\< R^2 \>$.
Our numerical results are consistent with these predictions.

We began by analysing the moments $\<r^{2k}\>_N$
using the method of differential approximants \cite{Gu87,Guttmann_89},
with the aim of confirming the predicted leading exponents $2k\nu = 3k/2$
and $q_k = 3k/2 - 59/32$.
It is a previously observed feature
of the method of differential approximants (DA)
that its application to the analysis of SAW moment series
is less accurate than might be expected \cite{Gu87}.
For example, DA analysis of
a 27-term square-lattice $\<R_e^2\>_N$ series,
biased at a critical point of 1,
produces estimates of $2\nu$ in the range 1.495--1.497 \cite{Gu87},
whereas an {\em unbiased}\/ analysis of
a 27-term SAW series on the same lattice
yields exponent estimates of $\gamma = 1.34364 \pm 0.00088,$
which is rather more accurate, as well as more precise.
This behaviour is most likely connected to the fact that
the method of differential approximants tacitly assumes
that the function is well approximated by a differentially finite
(D-finite) function,
i.e.\ the solution of a linear ordinary differential equation
with polynomial coefficients~\cite{G00}.
While there is strong evidence that neither SAWs nor SAPs are
D-finite \cite{G00},
it nevertheless appears that the SAW and SAP counts are well approximated
by a D-finite function,
while the generating functions for SAW and SAP metric properties
(such as $\<R_e^2\>_N$) appear not to be.
Evidence for this remark includes the telling fact that
most of the differential approximants for $\<R^2\>$ (for
both SAWs and SAPs) are {\em defective}~\cite{Gu87},
which is usually a signal that the function being approximated
is not of the type tacitly assumed by the analysis.
For this reason, our DA analysis gives only
moderately accurate estimates of the leading exponents, both
ferromagnetic and antiferromagnetic, but no reliable information
as to the value of the subleading exponents.

Our DA analysis confirmed the expected leading behaviour
$\<r^{2k}\>_N \propto N^{3k/2}$,
the exponents being identified as
$1.4997(5),$ $ 2.998(6),$ $ 4.496(9),$ $ 5.996(12),$ and $ 7.496(12)$
for $k = 1,2,3,4,5$ on the square lattice, and
$1.4997(5),$ $2.996(7),$ $ 4.495(9),$ $ 5.997(12),$ and $7.500(10)$
for $k = 1,2,3,4,5$ on the triangular lattice.
DA analysis also gave reasonable estimates of the
leading antiferromagnetic exponent on the square lattice:
we found $q_k = 3k/2 - 59/32$ to an accuracy of
approximately 0.01, 0.04, 0.05, 0.05, 0.06
for $k = 1,2,3,4,5$, respectively.
It is possible that higher moments ($k > 1$) may have
non-analytic correction-to-scaling terms with exponent $\Delta_1 < 1.5$
which would then be  more prominent than the leading 
correction-to-scaling term of the second-moment quantities $\<R^2\>$.
A more prominent such singularity would also explain the relatively poor accuracy
of the DA analysis.
We allow for this possibility in our analysis, described immediately below,
but find no evidence for such a term.

We next proceeded to fit $\<r^{2k}\>_N$ to the asymptotic forms
\reff{rrt_apriori} and \reff{rrs_apriori},
setting $\nu=3/2$ and $q_k = 3k/2 - 59/32$
and investigating the quality of the fit for
a variety of possible values of $\Delta_1$ and $\Delta_1^{\rm AF}$.
Among other things, we considered the possibility that $\Delta_1 < 1$,
even though no such term is observed in the metric quantities $\<R^2\>_N$.

We first fitted the available series to the above forms with $\Delta_1 = 1/2$.
Estimates of the associated amplitude were, in all cases,
monotonically decreasing toward zero. Furthermore, as we increased the
number of subdominant terms included in the fit,
this amplitude approached zero more and more closely.
The data are insufficient to judge whether the rate of approach
to zero increased, but the entries were numerically smaller.
Also, estimates of the leading amplitude $c_{0,k}$ did not display the
sort of convergence we found in the analysis of $\<R^2\>_N$;
rather, the convergence {\em deteriorated}\/
as we increased the number of subdominant terms included in the fit.
Both of these observations suggest that there is no term
$c_{2,k}/N^{\Delta_1}$ with $\Delta_1 \approx 1/2$.
Similar behaviour was observed with
   $\Delta_1 = 11/16$, though we cannot say whether the effect was stronger
or weaker. That is to say, the analysis is insensitive to this level of
exponent change for these series.
This is consistent with the situation found in the analysis of $\<R^2\>$.
We conclude that there is no evidence of a correction term
$c_{2,k}/N^{\Delta_1}$ with $\Delta_1 < 1$.

We then reanalysed the data assuming that 
the only correction-to-scaling term,
other than integer powers of $1/N$,
was that with exponent $\Delta_1 = 3/2$,
exactly as found for the second-moment series.
As for the second-moment series, we retained only
analytic correction terms to the antiferromagnetic singularity.
As we increased the order of the fit, the leading amplitudes $c_{0,k}$ and
$d_{0,k}$ displayed improved convergence. This is usually an indicator that
the guessed asymptotic form is correct.
The higher-order amplitudes displayed less convincing convergence, but
we ascribe this to a lack of adequate data.
For the second moment, we have a 59-term series,
which converged rather well, as can be seen from Table \ref{tableR2es}.
But the convergence is much less impressive after only 32 terms---which
is all we have available for the higher moments.

Taken together, our results favour the most obvious conjecture,
which is that the subdominant behaviour is characterised by the same exponent
set as is observed for $\<R^2\>_N$.

In order to test the conclusion that the leading correction term
in all these series is the $1/N$ term,
we used the method of differential approximants on
a modified series obtained by subtracting the estimated
leading-order term from the original series:
that is, we analysed $\<r^{2k} \>_N - c_{0,k} N^{2k\nu}$.
We found that the series coefficients
behave like $N^{2k\nu - 1}$,
consistent with the conclusion that the leading correction term is $1/N$
and that the non-analytic correction-to-scaling term(s),
have exponent $\Delta_1 > 1$, consistent with our view that $\Delta_1 = 3/2$.

With the foregoing observations in mind, we obtained the following estimates
for the corresponding amplitudes for the square-lattice moments $\<r^{2k}\>_N$,
where we have
assumed a single correction-to-scaling exponent $\Delta_1 = 3/2$ associated
with the ferromagnetic singularity, and otherwise only analytic corrections
to both the ferromagnetic and antiferromagnetic singularities:
\begin{eqnarray}
  \hbox{$k=2$:} &  &
c_{0,2}=0.860(2),\,  c_{1,2}=1.9(2),
      \nonumber \\
& &
d_{0,2}=0.139(5),\, d_{1,2} = -0.03(2), \\[4mm]
  \hbox{$k=3$:} &  &
c_{0,3}=1.184(5),\,  c_{1,3}=3(1),
      \nonumber \\
& &
d_{0,3}=0.193(5),\,  d_{1,3}=-2(1), \\[4mm]
  \hbox{$k=4$:} &  &
c_{0,4}=1.907(10),\,  c_{1,4}=2.5(5),
      \nonumber \\
& &
d_{0,4}=0.310(3),\, d_{1,4} = -0.46(5),\\[4mm]
  \hbox{$k=5$:} &  &
c_{0,5}=3.434(10),\,  c_{1,5}=-3(1),
      \nonumber \\
& &
d_{0,5}=0.551(3),\, d_{1,5} = -1.6(2).
\end{eqnarray}
As a check we verify relation \reff{rel-a0d0}. Our results for $c_N$ predict
$d_{0,k}/c_{0,k} = 0.161292(2)$, a relation that is well satisfied by our 
results for $c_{0,k}$ and $d_{0,k}$.

We can now provide a direct estimate of the invariant ratios
$M_{2k,N} = {\< r^{2k} \>_N / \< r^2 \>_N^k}$
in the limit $N \to\infty$.
{}From the above amplitude estimates, we have, for the square lattice,
\begin{eqnarray}
M_{4,\infty} &=& 1.446(3) \\
M_{6,\infty} &=& 2.581(11) \\
M_{8,\infty} &=& 5.391(28) \\
M_{10,\infty} &=& 12.59(4).
\end{eqnarray}
These estimates agree well with the estimates \reff{M4CPRV}--\reff{M10CPRV}
obtained previously \cite{Caracciolo-etal-distributionf},
but are a factor 5--10 less precise.

For the triangular lattice,
there is of course no ``antiferromagnetic'' singularity,
so that the terms corresponding to the amplitudes $d_{j,k}$ are absent.
We find from the triangular lattice data:
\begin{eqnarray}
\hbox{$k=2$:} &  &
c_{0,2}=0.7330(9),\,  c_{1,2}=1.2(2),\, c_{2,2}=-5(1),	\\[4mm]
\hbox{$k=3$:} &  &
c_{0,3}=0.934(2),\, c_{1,3} = 1.5(5),  \\[4mm]
\hbox{$k=4$:} &  &
c_{0,4}=1.383(3),\, c_{1,4} =2(1), \\[4mm]
\hbox{$k=5$:} &  &
c_{0,5}=2.31(3),\, c_{1,5} =-8.4(5).
\end{eqnarray}
{}From the above amplitude estimates, we have, for the triangular lattice,
\begin{eqnarray}
M_{4,\infty} &=& 1.446(2) \\
M_{6,\infty} &=& 2.588(5) \\
M_{8,\infty} &=& 5.381(12) \\
M_{10,\infty} &=& 12.64(15).
\end{eqnarray}
These estimates agree well with those found for the square lattice,
confirming the expected universality.
Therefore they are also in agreement with
the field-theory estimates \reff{M4CPRV}--\reff{M10CPRV}, though less precise.

\subsection{Non-Euclidean-invariant moments of the distribution function}
   \label{sec3.6}

In this section we discuss the behaviour of the
non-rotationally-invariant moments:
$\<r^{2k} \cos 4\theta \>_N$ with $k=2,3,4$ and
$\<r^{8} \cos 8\theta \>_N$ for the square lattice, and
$\<r^{2k} \cos 6\theta \>_N$ with $k=3,4$ for the triangular lattice.

Let us first consider the triangular lattice.
We began by analysing the series
using the method of differential approximants, with the aim of
determining the leading exponent.
For the triangular lattice we write
\begin{eqnarray}\label{rnrst}
\<r^{2k} \cos 6\theta \>_N
 & \sim &N^{2k\nu -\Delta_{\rm nr}}[a_{0,k}
+ a_{1,k}/N^{\Delta} +  a_{2,k}/N +\cdots ]\; .
\end{eqnarray}
Using first- and second-order differential approximants,
we found $\Delta_{\rm nr} = 3.00 \pm 0.10$ for $k=3$
and $\Delta_{\rm nr} = 2.95 \pm 0.10$ for $k=4$,
from which we conjecture that
$\Delta_{\rm nr} = 4\nu = 3$ exactly, as predicted in Section~\ref{sec2.1}.

Fitting the triangular-lattice data to the asymptotic form~\reff{rnrst},
we found good convergence only if we set the correction-to-scaling exponent
to a value $\Delta \approx 1/2$---in
sharp contrast to situation for the corresponding
rotationally invariant moments, where we found $\Delta = 3/2$.
To test the conjecture that the leading correction is $N^{-1/2}$,
we subtracted the estimated leading term $a_{0,k}N^{2k\nu -\Delta_{\rm nr}}$
from $\<r^{2k} \cos 6\theta \>_N$ and analysed the resulting series.
It was found to behave as
$a_{1,k}N^{2k\nu -\Delta_{\rm nr} - 0.50 \pm 0.10}$,
implying that $\Delta = 0.5 \pm 0.10$.
At this stage, we have no theoretical explanation for this
numerical observation.
Setting $\Delta_{\rm nr} = 3$ and $\Delta = 1/2$
and assuming subsequent half-integer terms in the asymptotic
expansion \reff{rnrst}, we obtained the following estimates for the
triangular-lattice amplitudes:
\begin{eqnarray}
\hbox{$k=3$:} &  &
a_{0,3}=1.120(3),\, a_{1,3} = -1.95(5),\, a_{2,3}=1.7(4)   \\[4mm]
\hbox{$k=4$:} &  &
a_{0,4}=4.05(5),\, a_{1,4} = -9(1),\, a_{2,4}=20(4).
\end{eqnarray}

For the square lattice, equation \reff{rnrst} needs to be modified by the
addition of a term representing the antiferromagnetic singularity, so we write
\begin{eqnarray}\label{rnrs}
\<r^{2k} \cos 4\theta \>_N
 & \sim &N^{2k\nu -\Delta_{\rm nr}}[a_{0,k}
+ a_{1,k}/N^{\Delta} +  a_{2,k}/N + \cdots ] \nonumber \\
& &  +\;  (-1)^N N^{\psi}[b_{0,k} + b_{1,k}/N^1
 + b_2/N^2 + \cdots].
\end{eqnarray}
{}From first- and second-order differential approximants applied to
the square-lattice data, we found
$\Delta_{\rm nr} = 1.46 \pm 0.03$,
$\Delta_{\rm nr} = 1.45 \pm 0.06$
and $\Delta_{\rm nr} = 1.44 \pm 0.09$
for $k = 2,3,4$, respectively.
{}From these results we conjecture that
$\Delta_{\rm nr} = 2\nu = 3/2$ exactly,
as predicted in Section~\ref{sec2.1}.
Differential-approximant analysis also gave reasonable estimates of
the the leading antiferromagnetic exponent: 
we found
$\psi = 2k\nu-\Delta_{\rm nr}-3+\gamma = 2k\nu - 101/32$,
accurate to $\pm 0.013$ for $k=2$, $\pm 0.05$ for $k=3$,
and $\pm 0.15$ for $k=4$.
This expression for $\psi$ is different from the one which 
is the natural generalization of the result for the Euclidean-invariant 
moments, $2k\nu-\Delta_{\rm nr} + \alpha - 1 - \gamma = 2k\nu - 107/32$, which
is excluded from the analysis: the difference is $6/32 = 0.1875$,
much larger than the errors.

Fitting the data to the above asymptotic form \reff{rnrs}
with $\Delta_{\rm nr} = 3/2$ and $\psi = 2k\nu - 101/32$,
and assuming only analytic corrections to scaling
at the antiferromagnetic critical point, as found for all the other series,
we again found good convergence only if we set
the ferromagnetic correction-to-scaling exponent $\Delta$
to approximately $1/2$,
just as was found in the analysis of the triangular-lattice data.
As in the triangular-lattice analysis,
we verified this conjecture by subtracting the estimated leading term
$a_{0,k}N^{2k\nu -\Delta_{\rm nr}}$
from $\<r^{2k} \cos 4\theta \>_N$ and analysing the resulting series, which
was found to behave as $a_{1,k}N^{2k\nu -\Delta_{\rm nr} - 0.497 \pm 0.005}$.
This is strong support for an $N^{-1/2}$ correction.

Setting $\Delta_{\rm nr} = 3/2$, $\psi = 2k\nu - 101/32$ and
$\Delta = 1/2$ and assuming subsequent half-integer terms in the asymptotic
expansion of the ferromagnetic singularity
\reff{rnrs}, we obtained the following estimates for the
square-lattice amplitudes:
\begin{eqnarray}
\hbox{$k=2$:} &  &
a_{0,2}=1.148(6),\, a_{1,2} = -1.70(5),\, a_{2,2}=2.7(3),
    \nonumber \\
  & &  b_{0,2}=0.060(5),\, b_{1,2} = 0.6(2), \\[4mm]
\hbox{$k=3$:} &  &
a_{0,3}=3.200(10),\, a_{1,3} = -6.25(10),\, a_{2,3}=12.0(5),
    \nonumber \\
  & &  b_{0,3}=0.175(10),\, b_{1,3} = 1.2(4),   \\[4mm]
\hbox{$k=4$:} &  &
a_{0,4}=8.90(10),\, a_{1,4} = -20(2),\, a_{2,4}=50(8),
    \nonumber \\
  & &  b_{0,4}=0.47(5),\, b_{1,4} = 3(1).
\end{eqnarray}

Finally, we analyzed the series $\<r^8 \cos 8\theta \>_N$
on the square lattice.
A differential-approximant analysis gave
$\<r^8 \cos 8\theta \>_N \propto N^{3.07 \pm 0.1}$.
We conjecture that the exponent is exactly 3,
consistent with the behaviour
$\<r^8 \cos 8\theta \>_N \propto N^{8\nu - \Delta_{{\rm nr},8}}$
with $\Delta_{{\rm nr},8} = 3$.
This is exact at the mean-field level and also for the
two-dimensional Ising model. In the same spirit that we
previously conjectured that $\Delta_{\rm nr}=2\nu$,
we now conjecture that $\Delta_{{\rm nr},8} = 4\nu$.
The antiferromagnetic exponent in $\<r^8 \cos 8\theta \>_N$
was estimated to be $\psi = 1.35 \pm 0.05$, which, by analogy with the
antiferromagnetic exponents for $\<r^{2k} \cos 4\theta \>_N $, 
we conjecture is exactly $8 \nu - \Delta_{{\rm nr},8} - 3 + \gamma = 43/32.$
We found the
subsequent analysis consistent with only analytic corrections at the
antiferromagnetic critical point.
At the ferromagnetic critical point, the data were again consistent with a
leading $N^{-1/2}$ correction-to-scaling term.
In an identical notation to that used above, we find the
amplitudes to be:
\begin{eqnarray}
 &  &
a_{0}=135(2),\, a_{1} = -540(10),\, a_{2}=1440 (50),
    \nonumber \\
  & &  b_{0}=6.5(1),\, b_{1} = -11(3).
\end{eqnarray}

\section{Monte Carlo analysis}  \label{sec4}

\subsection{Summary of our data}  \label{sec4.1}

We have also generated extensive Monte Carlo data, using the pivot algorithm
\cite{Lal,MacDonald,Madras-Sokal,Li-Madras-Sokal},
for SAWs on the square lattice at
selected values of $N$ in the range $40 \le N \le 4000$,
measuring the observables $\<R_e^2\>$, $\<R_g^2\>$ and $\<R_m^2\>$.
The results are collected in Tables \ref{tableMC1} and \ref{tableMC2}.
Unfortunately, in some runs we did not measure all observables:
in particular, for larger values of $N$, the statistics available for
$\<R^2_m\>$ are much smaller than for the other radii.
The statistics range from $10^9$ to $10^{10}$ pivot iterations per point,
or approximately $10^7$ to $10^9$ times the integrated autocorrelation time
$\tau_{{\rm int},A}$ for these observables.\footnote{High-precision Monte Carlo
data for $\<R_e^2\>_N$ have been kindly provided by Peter Grassberger. 
His results have been merged with ours and appear in Tables 
\ref{tableMC1} and \ref{tableMC2}.}

We begin by analyzing the Monte Carlo data in an unbiased way,
in order to extract the leading amplitudes and the
correction-to-scaling exponents and amplitudes (Section~\ref{sec4.2}).
Then we compare the Monte Carlo data,
which lie at relatively high $N$ but are afflicted by statistical errors,
with the formulae \reff{sqfita}--\reff{sqfitc} obtained by extrapolating
the series data, which lie at much smaller $N$ but are exact
(Section~\ref{sec4.3}).

\subsection{Data analysis}   \label{sec4.2}

In order to determine the leading amplitudes and the
correction-to-scaling exponent(s) and amplitude(s),
we have analysed the three quantities
$\<R_e^2\>$, $\<R_g^2\>$ and $\<R_m^2\>$.
We first tried nonlinear least-squares fits of the form\footnote{
   Note that we have rescaled the length $N$ by a fixed parameter $N_0$
   that has always been taken equal to $N_0=750$.
   The purpose of this rescaling is to minimize the covariance
   between the estimates of $b$ and $\Delta$.
   (The optimal choice is to take $N_0$ equal to the weighted geometric mean
    of the $N$ values occurring as data points.)
   As a consequence, the relative error on $b$ is a factor of 3--4 smaller
   than in fits with $N_0=1$.
   The error on $\Delta$ does not depend on the choice of $N_0$.
}
\be
\< R^2 \>_N/N^{2\nu}  \;=\; a + b (N/N_0)^{-\Delta}   \;,
\label{fitvar}
\ee
in which $\nu$ has been fixed equal to $3/4$
and the parameters $a,b,\Delta$ are free.
In these fits we use only the data with
$N \ge$ some cutoff value $N_{\rm min}$;
we then vary $N_{\rm min}$ systematically
and investigate the quality of the fit
(see Tables~\ref{tableR2evar}, \ref{tableR2gvar} and \ref{tableR2mvar}).
For $N_{\rm min} \gtapprox 60$ the $\chi^2$ values are reasonable.
The fits are stable and the error bars reasonable up to
$N_{\rm min} \approx 700$;
after this, the error bars increase drastically and the estimates
should be considered unreliable.

Let us consider first $\<R^2_e\>$ and $\<R^2_g\>$,
the radii for which we have the best statistics.
The fit for the radius of gyration is extremely stable\footnote{
   The $\chi^2$ of the fit is somewhat large, and the corresponding
   confidence level too small.
   Since the confidence level actually gets {\em worse}\/
   as $N_{\rm min}$ grows,
   the cause does not seem to be corrections to scaling.
   The most likely interpretation is that our error bars are,
   for some unknown reason,
   somewhat underestimated for large values of $N$.
   (Perhaps we have underestimated the integrated autocorrelation time
    by failing to include a sufficient amount of the tail of the
    autocorrelation function.)
}
and clearly suggests $\Delta = 1$.
No subleading exponent with $\Delta < 1$ appears to be present;
in particular, this excludes $\Delta = 11/16 =  0.6875$
unless the corresponding amplitude is extremely small.
By contrast, the fit of $\<R_e^2\>$ gives estimates that vary with $N_{\rm min}$:
the estimate of $\Delta$ at first increases with $N_{\rm min}$,
then flattens off at $\Delta \approx 0.9$.
The theoretical prediction $\Delta = 11/16$  is again excluded,
but in this case a non-analytic correction $\Delta_1 < 1$
is still possible {\em a priori}\/.
However, we believe that a subleading exponent $\Delta_1 \approx 0.9$
is unlikely:
after all, it does not agree with any theoretical prediction;
and the observed behaviour can be explained equally well, as noted earlier,
by the competition between two correction terms of opposite sign,
provided that both terms are still sizable
in the range of $N$ that we are considering.
The fact that a range $500\le N \le 4000$ is insufficient to
see clearly that $\Delta = 1$ shows how difficult is the determination
of $\Delta$ and explains the wide range of contradictory results found
in previous work.

We can also consider $\<R^2_m\>$. The behaviour is similar to that observed for
$\<R^2_e\>$, although the errors are larger. Again, the data are compatible
with $\Delta \approx 0.9$, but, as before, we believe
that what we are observing is simply an effective exponent
arising from the competition between two correction terms of opposite sign.
Moreover, we are here probing shorter walks than in the case of $\<R^2_e\>$,
because of the lack of statistics for larger $N$:
the fit is effectively dominated by the data in the range
$300 \ltapprox N \ltapprox 1000$.

Since we have little evidence that $\Delta_1 = 11/16$,
the most likely possibility is that $\Delta_1 = 3/2$.
We have therefore checked whether our data can be fitted by the simple Ansatz
\be
\<R^2\>_N   \;=\; N^{2\nu} (a + b N^{-1} + c N^{-3/2}),
\label{fitfixed}
\ee
with $a$, $b$, $c$ free parameters. As before, we perform several
fits using only data with $N\ge N_{\rm min}$, varying
$N_{\rm min}$ systematically. The results are reported in
Tables \ref{tableR2efixed}, \ref{tableR2gfixed} and \ref{tableR2mfixed}.
The fit quality is good even for the lowest value of $N_{\rm min}$:
additional corrections do not play much role for $N\gtapprox 60$.
Notice that for $\<R^2_g\>$ the constant $c$ is very small, explaining
why the previous fit gave $\Delta = 1$ essentially without
corrections. For $\<R^2_e\>$ and $\<R^2_m\>$, $c$ is instead sizable and of
opposite sign with respect to $b$, giving in
fit \reff{fitvar} an effective exponent $\Delta < 1$.

The leading amplitudes are extremely stable and, using the
data with $N_{\rm min} = 200$,  we can estimate
\begin{eqnarray}
a_e &=&  0.77121 \pm 0.00004  \label{4.3} \\
b_e &=&  1.17 \pm 0.05  \\
c_e &=&  -2.8 \pm 0.7,  \\[4mm]
a_g &=&  0.108207 \pm 0.000007   \\
b_g &=&  0.108 \pm 0.010   \\
c_g &=&  0.0 \pm 0.2,  \\[4mm]
a_m &=&  0.33903 \pm 0.00004   \label{4.7} \\
b_m &=&  0.46 \pm 0.05 \\\label{4.8}
c_m &=&  -1.1 \pm 0.7  \; .
\end{eqnarray}
where the error bars are 68\% confidence limits.
We can compare these results with the series estimates
\reff{sqfita}--\reff{sqfitc},
and note that they agree well within quoted errors with one exception:
$a_g$ differs by three error bars
from one set of the corresponding series estimate.
We consider the stated errors of this series estimate to be
anomalously low, compared to all other error estimates,
and thus not to be taken literally.

We have also considered the universal amplitude ratios $A_N$ and $B_N$.
The most notable feature of the raw data for $A_N$ (Table~\ref{tableMC2})
is its nonmonotonicity:
at first $A_N$ decreases, reaching a minimum at $N \approx 130$;
then it increases.
This immediately suggests the presence of two correction-to-scaling terms
of opposite sign, in agreement with the analysis presented above.
We have therefore analysed the ratios $A_N$ and $B_N$
by performing a fit of the form
\be
\scro_N \;=\; a + b N^{-1} + c N^{-3/2}   \;.
\ee
The results are reported in Tables
\ref{tableAfixed} and \ref{tableBfixed}.
Again the quality of the fits is quite good, and we obtain the final
estimates (again we use conservatively the data for $N_{\rm min} = 200$)
\begin{eqnarray}
A &=& 0.140310 \pm 0.000011  \\
B &=& 0.439614 \pm 0.000050,
\end{eqnarray}
where the error bars are 68\% confidence limits.\footnote{
   Cardy and Mussardo \cite{Cardy-Mussardo} have used the
   form-factor method, applied to the exact $S$-matrix
   of the massive $O(n)$ model, to derive the estimates
   $A \approx 0.126$ and $B \approx 0.420$.
   This is impressive accuracy for a first-principles
   theoretical calculation;  the approximately 5\% error
   is about what one expects from the one-particle
   approximation used in this computation.
}

Next we analysed the CSCPS combination \reff{fN-def}.
Conformal-invariance theory \cite{Cardy-Saleur,CPS:Cardy-Saleur}
predicts that $\lim_{N\to\infty} F_N = 0$,
and we confirm this prediction numerically to very high precision:
see Table \ref{tableFfixed}.
Therefore, $f_N$ is expected to scale as $N^{2\nu-\delta}$
where $\delta$ is some subleading exponent.
Our data are consistent with $\delta=3/2$, although not so accurate
as to establish it unambiguously;
in other words, $f_N$ appears to approach a nonzero constant as $N\to\infty$.
This means, as noted earlier, that the $1/N$ correction is absent within
our errors.
A fit to a constant gives the results reported in Table~\ref{tableFRE}.
The results show an initial downward trend with
$N_{\rm min}$ and then increase again. A stable region may be identified
for $N_{\rm min} \ge 250$. For $N_{\rm min} = 250$ we have
\be
f  \;=\; 0.68 \pm 0.14\; .
\ee
This is in agreement with, but less precise than, the result
$f = 0.79 \pm 0.03 $ obtained from series analysis.

Finally, we have tried to see whether our estimated values of $A$ and $B$
are consistent with simple rational values
that satisfy the CSCPS formula.
If we require the denominators to be $\le 1000$,
then the only possible values anywhere near our estimates are
\be
   A = {23 \over 164} \approx 0.1402439,
   \qquad
   B = {40 \over 91} \approx 0.4395604
\ee
and
\be
   A = {91 \over 648} \approx 0.1404321,
   \qquad
   B = {95 \over 216} \approx 0.4398148.
\ee
Our data are, however, precise enough to clearly exclude both guesses.
We therefore conjecture that $A$ and $B$
do {\em not}\/ take simple rational values, even though
one particular linear combination of them does.

\subsection{Comparison to series-extrapolation predictions}   \label{sec4.3}


We can also directly compare our raw Monte Carlo data to the
extrapolation formulas \reff{sqfita}--\reff{sqfitc}.
For this purpose we compute
\be
\chi^2  \;=\;
 \sum_{\rm MC\,data} {(R^2_{MC} - R^2_{\rm series})^2\over \sigma^2_{MC}}\; ,
\ee
where $R^2_{MC}$ is the Monte Carlo estimate, $\sigma_{MC}$ the
corresponding error, and $R^2_{\rm series}$ the
prediction of the extrapolations \reff{sqfita}--\reff{sqfitc}.
For $\<R^2_e\>$ we find that \reff{sqfita} describes the numerical data
rather well.
Indeed, $\chi^2 = 29$ for 19 data points.
The small remaining discrepancy is mainly due to the error on the coefficients. 
Indeed, if we use for the ferromagnetic part the results of the 
Monte Carlo fit with $N_{\rm min} = 40$ reported in Table 
\ref{tableR2efixed}---these estimates are compatible with 
the exact-enumeration ones reported in \reff{sqfita}---the $\chi^2$ drops to 
15. Analogous discussion applies to $\<R^2_m\>$. If we use \reff{sqfitc}
we obtain $\chi^2 = 55$ with 15 data points. But again it is enough to replace
the ferromagnetic coefficients obtained using exact enumeration with those 
obtained in the Monte Carlo fit, see Table \ref{tableR2mfixed}---they are 
fully compatible---to have $\chi^2 = 19$ with 15 points.
The situation is worse for $\<R^2_g\>$. Using all data we obtain 
$\chi^2 = 80$ with 18 data points. Such a result does not improve 
significantly if we change the coefficients in \reff{sqfitb} within error bars. 
This is related to the fact we have already noticed that the leading 
amplitude for $\<R^2_g\>_N$ reported in \reff{sqfitb} significantly differs 
from the Monte Carlo estimate obtained for any value of $N_{\rm min}$.

\section{Conclusions and open questions}   \label{sec5}

In this study of the SAW correction-to-scaling exponents,
we have seen a consistent picture emerging, given independent support
by both Monte Carlo and series analysis.
We have presented compelling evidence
that the first non-analytic correction term
in the generating function for both SAWs and SAPs,
as well as in several Euclidean-invariant metric properties,
is $\Delta_1 = 3/2$, as predicted by Nienhuis
some 20 years ago \cite{Nienhuis_82,Nienhuis_84}. 
We find no evidence for the presence of an exponent $\Delta_1 = 11/16$
in SAWs and SAPs on the square and triangular lattices. 
Our analysis of the interplay between dominant and
subdominant correction-to-scaling terms also enables us to
explain quantitatively why many earlier analyses gave incorrect
conclusions, predicting exponents $\Delta_1 < 1$. 
For certain observables, we find pairs of correction terms 
of opposite sign that conspire to give effective exponents 
that are smaller than both of the individual exponents. Thus, corrections
behaving as $a/N + b/N^{3/2}$ with $ab < 0$
were incorrectly identified with a single correction term
$c/N^{\Delta_1}$ with $\Delta_1 <1$.

Monte Carlo and series analysis turn out to complement each other well.
Series provide a basis for calculating the amplitudes of several subdominant
asymptotic terms, while the Monte Carlo data frequently provide
greater accuracy for the estimate of the leading amplitudes.

We have also studied the asymptotic behaviour of several
non-Euclidean-invariant quantities.
Their leading behaviour is characterized by a new exponent
$\Delta_{\rm nr}$.
We find compelling evidence that $\Delta_{\rm nr} = 2 \nu$
on the square lattice and $\Delta_{\rm nr} = 4 \nu$
on the triangular lattice,
confirming the conjecture of \cite{CPRV-2dNle2,CPRV-2punti}. We also
computed the leading correction-to-scaling exponent in these observables,
finding $\Delta_1 \approx 0.5$.
We are unaware of any theoretical prediction for this quantity.

We have also determined the dominant and subdominant exponents
characterizing the ``antiferromagnetic singularity" of the square lattice.
These exponent predictions are for the most part new.

We also tested the CSCPS relation $\lim_{N\to\infty} F_N = 0$
[cf.\ \reff{CSC}], which arises from conformal field theory
\cite{Cardy-Saleur,CPS:Cardy-Saleur}.
Both our Monte Carlo and series work
are completely consistent with the CSCPS relation.
Further, we find that the $1/N$ correction term
in $F_N$ is absent, so that $F_N \sim {\rm const} \times N^{-\Delta_1}$
with $\Delta_1 = 3/2$.
The absence of this analytic correction-to-scaling term
implies a new amplitude relation \reff{eq_CSCPS_subdominant}.

Finally, we remark that our numerical estimates for the
universal amplitude ratios $A$ and $B$
are now so precise as to allow us to rule out the
possibility that these are rational numbers
with small integer denominators.
Some other universal amplitude ratios include powers of $\pi$
\cite{Cardy-Guttmann}, so it is possible
that $A$ and $B$ are combinations of $\pi$ and rational numbers;
but there is no {\em a priori}\/ reason why powers of $\pi$
should enter into the amplitude ratios $A$ and $B$.

\section*{Acknowledgments}

The authors wish to thank Giovanni Ferraro for many helpful discussions and
Raghu Varadhan for interesting suggestions.
We also wish to thank Debra Bennett-Wood and Bin Li for helping to generate
some of the series and Monte Carlo data, respectively;
Aleks Owczarek for helpful comments on the manuscript;
and Peter Grassberger for providing high-precision Monte Carlo data
that have been included in the analysis.
Finally, we wish to thank an anonymous referee for important comments
concerning the renormalization-group explanation of the cancellation
of the analytic corrections to the CSCPS relation.

This research was supported in part by
the Consiglio Nazionale delle Ricerche (S.C. and A.P.),
the Australian Research Council (A.J.G. and I.J.),
the U.S.\ National Science Foundation grants DMS--8911273, DMS--9200719,
PHY--9520978, PHY--9900769, PHY--0099393 and PHY--0424082 (A.D.S.),
the U.S.\ Department of Energy contract DE-FG02-90ER40581 (A.D.S.),
the NATO Collaborative Research grant CRG 910251 (S.C. and A.D.S.),
and a grant from the New York University Research Challenge Fund (A.D.S.).
Acknowledgment is also made to the donors of the Petroleum Research Fund,
administered by the American Chemical Society, for partial support
of this research under grants 21091--AC7 and 25553--AC7B--C.
A.N.R.\ and I.J.\ are pleased to acknowledge computational facilities
provided by both VPAC and APAC.

\clearpage

%
%
\begin{table}[p]
\hspace*{-1cm}
\tiny
\renewcommand{\arraystretch}{1.10}
\begin{tabular}{rrrrr}  \hline  \hline \\
 \normalsize $N$  & \normalsize $c_N$ &
 \normalsize $\frac14 c_N \langle R^2_e \rangle _N$ &
 \normalsize $\frac14 (N+1)^2 c_N \langle R^2_g \rangle _N$  &
 \normalsize $\frac14 (N+1) c_N \langle R^2_m \rangle _N$ \\ \\
\hline \hline
1  &  4  &  1  &  1  &  1  \\
2  &  12  &  8  &  14  &  11  \\
3  &  36  &  41  &  116  &  74  \\
4  &  100  &  176  &  722  &  390  \\
5  &  284  &  679  &  3887  &  1801  \\
\hline \hline
\end{tabular}
\vspace{1cm}
\caption{
   Exact enumeration data for SAWs on the square lattice.
}
\label{table1}
\end{table}

%
%
\begin{table}
\hspace*{-1cm}
\tiny
\renewcommand{\arraystretch}{1.10}
\begin{tabular}{rrrrr}  \hline  \hline \\
 \normalsize $N$  & \normalsize $c_N$ &
 \normalsize $\frac16 c_N \langle R^2_e \rangle _N$ &
 \normalsize $\frac16 (N+1)^2 c_N \langle R^2_g \rangle _N$  &
 \normalsize $\frac16 (N+1) c_N \langle R^2_m \rangle _N$ \\ \\
\hline \hline
1  &  6  &  1  &  1  &  1  \\
2  &  30  &  12  &  22  &  17  \\
3  &  138  &  97  &  282  &  178  \\
4  &  618  &  654  &  2778  &  1476  \\
5  &  2730  &  3977  &  23305  &  10667  \\
\hline \hline
\end{tabular}
\vspace{1cm}
\caption{
   Exact enumeration data for SAWs on the triangular lattice.
}
\label{table2}
\end{table}

%
%
\begin{table}[p]
\centering
\begin{tabular}{rrrrr}
\hline\hline
\\[-0.3cm]
\multicolumn{1}{c}{$N$}  &
  \multicolumn{1}{c}{${1 \over 4} c_N \< r^4 \>_N$}  &
  \multicolumn{1}{c}{${1 \over 4} c_N \< r^6 \>_N$}  &
  \multicolumn{1}{c}{${1 \over 4} c_N \< r^8 \>_N$}  &
  \\[0.2cm]
\hline\hline
   1&   1&   1&   1\\
   2&   24&   80&   288\\
   3&   233&   1481&   10313\\
   4&   1552&   15584&   171712\\
   5&   8261&   118741&   1876421\\
   6&   40128&   761824&   15997248\\
\hline\hline
\end{tabular}
\caption{
   Exact enumeration data for SAWs on the square lattice.
   Rotationally invariant moments.
}
\label{sq_ri_moments}
\end{table}

\begin{table}[p]
\centering
\begin{tabular}{rrrr}
\hline\hline
\\[-0.3cm]
\multicolumn{1}{c}{$N$}  &
  \multicolumn{1}{c}{${1 \over 4} c_N \< r^4 \cos 4 \theta\>_N$}  &
  \multicolumn{1}{c}{${1 \over 4} c_N \< r^6 \cos 4 \theta\>_N$}  &
  \multicolumn{1}{c}{${1 \over 4} c_N \< r^8 \cos 4 \theta\>_N$}  \\[0.2cm]
\hline\hline
1& 1 &                   1 &            1\\
2& 8 &                   48 &           224\\
3& 41 &                  521 &          5513\\
4& 176 &                 3616 &         64832\\
5& 517 &                 18581 &        515461\\
6& 2464 &                96672 &        3460288\\
\hline\hline
\end{tabular}
\caption{
   Exact enumeration data for SAWs on the square lattice.
   Non-rotationally-invariant moments.
}
\label{sq_nri_moments}
\end{table}

\begin{table}[p]
\centering
\begin{tabular}{rrrr}
\hline\hline
\\[-0.3cm]
\multicolumn{1}{c}{$N$}  &
  \multicolumn{1}{c}{${1 \over 6} c_N \< r^4 \>_N$}  &
  \multicolumn{1}{c}{${1 \over 6} c_N \< r^6 \>_N$}  &
  \multicolumn{1}{c}{${1 \over 6} c_N \< r^8 \>_N$}  \\[0.2cm]
\hline\hline
  1&    1&  1&   1\\
    2&  36&   120&  420\\
    3&  529&   3337&  22993\\
    4&  5454&   53358&  579918\\
    5&  46169&   633185&  9725849\\
    6&  344428&    6221884&  126456796\\
    7&  2352769&   53647549&  1380311377\\
    8&  15060090&   420194610&  13256099610\\
    9&  91701871&   3057404227&  115436446639\\
   10&  536695548&  20985811596&  930424151244\\
   11&  3041620465&  137386509145&  7043622904369\\
   12&  16784388968&   864860517248&  50628426215432\\
   13&  90564392107&  5267777416675&  348381660817291\\
   14&  479388030946&  31195330864090&  2309752803978322\\
   15&  2495911050345&  180302936975925&  14829889785198921\\
   16&  12808562012852&   1020273312831596&  92588968138883348\\
   17&  64901247920059&  5666785971562159&  564018088195524619\\
   18&  325170745810666&  30958613694252346&  3361656410695492858\\
   19&  1612871078099977&   166655904211475269&  19649810153900928217\\
   20&  7927964773508104&   885333404431705216&  112867935343523712424\\
   21&  38652473796950531&   4647266177110051355&  638156051853903239891\\
   22&  187056802703356296&    24130876287242419704& 3556820174217345377400\\
\hline\hline
\end{tabular}
\caption{
   Exact enumeration data for SAWs on the triangular lattice.
   Rotationally invariant moments.
}
\label{tri_ri_moments}
\end{table}

\clearpage

\begin{table}[p]
\centering
\begin{tabular}{rrr}
\hline\hline
\\[-0.3cm]
\multicolumn{1}{c}{$N$}  &
  \multicolumn{1}{c}{${1 \over 6} c_N \< r^6 \cos 6\theta \>_N$}  &
  \multicolumn{1}{c}{${1 \over 6} c_N \< r^8 \cos 6\theta \>_N$}  \\[0.2cm]
\hline\hline
   1&    1&    1\\
    2&    12&    96\\
    3&    97&    1609\\
    4&    654&    17454\\
    5&    3977&    151649\\
    6&    22684&    1149148\\
    7&    123721&    7935829\\
    8&    652842&    51236610\\
    9&    3357439&    314319571\\
   10&    16914348&    1852261068\\
   11&    83777857&    10566164665\\
   12&    409089560&    58677117008\\
   13&    1973505067&    318573263587\\
   14&    9421326322&    1696583222746\\
   15&    44567944521&    8885936605365\\
   16&    209144745044&    45868003547852\\
   17&    974497840243&    233746526628199\\
   18&    4511869867210&    1177691322037546\\
   19&    20770914530257&    5873419831448317\\
   20&    95130303643048&    29024633960838784\\
   21&	  433664585252891&   142245181072370291\\
   22&	  1968525488778840&  691879404495232056\\

\hline\hline
\end{tabular}
\caption{
   Exact enumeration data for SAWs on the triangular lattice.
   Non-rotationally-invariant moments.
}
\label{tri_nri_moments}
\end{table}

\begin{table}[p]
\centering
\tiny
\hspace*{-30pt}
\begin{tabular}{rrrrr}
\hline\hline
\multicolumn{1}{c}{$N$} &
  \multicolumn{1}{c}{$a_{0}$}  &
  \multicolumn{1}{c}{$a_{1}$}  &
  \multicolumn{1}{c}{$a_{2}$}  &
  \multicolumn{1}{c}{$a_{3}$}\\
\hline
19 &  0.712401 & 0.847374 &$-$1.932382 & 1.872489 \\
 20 &  0.712338 & 0.854132 &$-$1.970591 & 1.933229 \\
 21 &  0.712262 & 0.862841 &$-$2.021181 & 2.015860 \\
 22 &  0.712207 & 0.869360 &$-$2.060036 & 2.080978 \\
 23 &  0.712154 & 0.876015 &$-$2.100682 & 2.150783 \\
 24 &  0.712108 & 0.882131 &$-$2.138917 & 2.217997 \\
 25 &  0.712067 & 0.887809 &$-$2.175211 & 2.283237 \\
\hline\hline
\end{tabular}

\caption{
   Fit to $\<R^2\> = N^{3/2} (a_{0} + a_{1} N^{-1}  + a_{2} N^{-3/2} +
           a_{3} N^{-2} )$
   for SAWs on the triangular lattice.
   Data is for $\<R^2_e\>_N$ at top, then $\<R^2_g\>_N$ at middle,
   then $\<R^2_m\>_N$ at bottom.
   The fit for $\<R^2_m\>_N$ includes only terms up to order $(N^{-3/2})$.
}
\label{tableR2t}
\end{table}

\begin{table}[p]
\centering
\hspace*{-30pt}
\begin{tabular}{rrrrrrrrr}
\hline\hline
\multicolumn{1}{c}{$N$} &
  \multicolumn{1}{c}{$a_{0,e}$}  &
  \multicolumn{1}{c}{$a_{1,e}$}  &
  \multicolumn{1}{c}{$a_{2,e}$}  &
  \multicolumn{1}{c}{$a_{3,e}$}  &
  \multicolumn{1}{c}{$a_{4,e}$}  &
  \multicolumn{1}{c}{$b_{0,e}$}  &
  \multicolumn{1}{c}{$b_{1,e}$}  &
  \multicolumn{1}{c}{$b_{2,e}$}  \\
\hline

 33 &  0.771324 & 1.127229 &$-$2.737952 & 3.927648 &$-$2.772841 & 0.124654 &$-$0.037820 & 0.655979 \\
 34 &  0.771291 & 1.137155 &$-$2.846606 & 4.373539 &$-$3.423234 & 0.124636 &$-$0.036728 & 0.639632 \\
 35 &  0.771305 & 1.132833 &$-$2.798505 & 4.172870 &$-$3.125666 & 0.124628 &$-$0.036255 & 0.632312 \\
 36 &  0.771288 & 1.138263 &$-$2.859903 & 4.433123 &$-$3.517795 & 0.124619 &$-$0.035663 & 0.622865 \\
 37 &  0.771293 & 1.136612 &$-$2.840943 & 4.351507 &$-$3.392907 & 0.124616 &$-$0.035484 & 0.619916 \\
 38 &  0.771279 & 1.141331 &$-$2.895949 & 4.591870 &$-$3.766261 & 0.124609 &$-$0.034975 & 0.611270 \\
 39 &  0.771286 & 1.138877 &$-$2.866927 & 4.463195 &$-$3.563461 & 0.124605 &$-$0.034711 & 0.606663 \\
\hline\hline
\end{tabular}

\caption{
   Fit $\<R^2_e\>_N = N^{3/2} (a_{0,e} + a_{1,e} N^{-1}  + a_{2,e} N^{-3/2} +
           a_{3,e} N^{-2} + a_{4,e} N^{-5/2}) + (-1)^N N^{-11/32} (b_{0,e} + b_{1,e}/N +
           b_{2,e}/N2)$
   for SAWs on the square lattice.
}
\label{tableR2es}
\end{table}

\begin{table}[p]
\centering
\hspace*{-30pt}
\begin{tabular}{rrrrrr}
\hline\hline
\multicolumn{1}{c}{$N$} &
  \multicolumn{1}{c}{$a_{0,g}$}  &
  \multicolumn{1}{c}{$a_{1,g}$}  &
  \multicolumn{1}{c}{$a_{2,g}$}  &
  \multicolumn{1}{c}{$b_{0,g}$}  &
  \multicolumn{1}{c}{$b_{1,g}$} \\ 
\hline
 33 &  0.108213 & 0.103702 & 0.100792 & 0.008333 & 0.004174 \\
 34 &  0.108216 & 0.103450 & 0.101736 & 0.008370 & 0.003016 \\
 35 &  0.108218 & 0.103242 & 0.102526 & 0.008339 & 0.003994 \\
 36 &  0.108220 & 0.103040 & 0.103306 & 0.008369 & 0.003017 \\
 37 &  0.108221 & 0.102873 & 0.103959 & 0.008345 & 0.003842 \\
 38 &  0.108223 & 0.102712 & 0.104599 & 0.008368 & 0.003025 \\
 39 &  0.108224 & 0.102579 & 0.105134 & 0.008349 & 0.003715 \\
 40 &  0.108225 & 0.102452 & 0.105653 & 0.008367 & 0.003039 \\
 41 &  0.108226 & 0.102347 & 0.106085 & 0.008352 & 0.003607 \\
 42 &  0.108227 & 0.102249 & 0.106500 & 0.008366 & 0.003058 \\
 43 &  0.108228 & 0.102168 & 0.106842 & 0.008355 & 0.003515 \\
 44 &  0.108228 & 0.102092 & 0.107166 & 0.008365 & 0.003078 \\
 45 &  0.108229 & 0.102032 & 0.107429 & 0.008357 & 0.003436 \\
 46 &  0.108229 & 0.101976 & 0.107674 & 0.008365 & 0.003100 \\
 47 &  0.108230 & 0.101933 & 0.107867 & 0.008359 & 0.003367 \\
 48 &  0.108230 & 0.101894 & 0.108043 & 0.008364 & 0.003122 \\
 49 &  0.108230 & 0.101865 & 0.108175 & 0.008360 & 0.003307 \\
 50 &  0.108230 & 0.101840 & 0.108290 & 0.008363 & 0.003145 \\
 51 &  0.108230 & 0.101823 & 0.108367 & 0.008361 & 0.003254 \\
 52 &  0.108230 & 0.101810 & 0.108428 & 0.008363 & 0.003166 \\
 53 &  0.108230 & 0.101804 & 0.108456 & 0.008362 & 0.003207 \\
 54 &  0.108230 & 0.101801 & 0.108470 & 0.008363 & 0.003187 \\
 55 &  0.108230 & 0.101804 & 0.108455 & 0.008363 & 0.003165 \\
 56 &  0.108230 & 0.101810 & 0.108427 & 0.008362 & 0.003207 \\
 57 &  0.108230 & 0.101821 & 0.108373 & 0.008364 & 0.003128 \\
 58 &  0.108230 & 0.101834 & 0.108307 & 0.008362 & 0.003226 \\
 59 &  0.108230 & 0.101852 & 0.108219 & 0.008364 & 0.003094 \\
\hline\hline
\end{tabular}

\caption{
     Fit $\<R^2_g\>_N = N^{3/2} (a_{0,g} + a_{1,g} N^{-1}  + a_{2,g} N^{-3/2})
     + (-1)^N N^{-11/32} (b_{0,g} + b_{1,g}/N )$
     for SAWs on the square lattice.
}
\label{tableR2gs}
\end{table}

\begin{table}[p]
\centering
\hspace*{-30pt}
\begin{tabular}{rrrrrrr}
\hline\hline
\multicolumn{1}{c}{$N$} &
  \multicolumn{1}{c}{$a_{0,m}$}  &
  \multicolumn{1}{c}{$a_{1,m}$}  &
  \multicolumn{1}{c}{$a_{2,m}$}  &
  \multicolumn{1}{c}{$a_{3,m}$}  &
  \multicolumn{1}{c}{$b_{0,m}$}  &
  \multicolumn{1}{c}{$b_{1,m}$}  \\
\hline
 33 &  0.339223 & 0.401114 &$-$0.849196 & 1.011789 & 0.036366 & 0.020608\\
 34 &  0.339252 & 0.395837 &$-$0.810037 & 0.930082 & 0.036379 & 0.020222 \\
 35 &  0.339210 & 0.403819 &$-$0.870217 & 1.057665 & 0.036397 & 0.019642 \\
 36 &  0.339233 & 0.399343 &$-$0.835945 & 0.983880 & 0.036407 & 0.019318 \\
 37 &  0.339198 & 0.406364 &$-$0.890517 & 1.103143 & 0.036422 & 0.018812 \\
 38 &  0.339216 & 0.402541 &$-$0.860364 & 1.036281 & 0.036430 & 0.018538 \\
 39 &  0.339188 & 0.408723 &$-$0.909807 & 1.147474 & 0.036442 & 0.018096 \\
\hline\hline
\end{tabular}

\caption{
   Fit $\<R^2_m\>_N = N^{3/2} (a_{0,m} + a_{1,m} N^{-1}  + a_{2,m} N^{-3/2} +
            a_{3,m} N^{-2}) + (-1)^N N^{-11/32} (b_{0,m} + b_{1,m}/N )$
   for SAWs on the square lattice.
}
\label{tableR2ms}
\end{table}

\begin{table}[p]
\centering
\hspace*{-30pt}
\begin{tabular}{rr@{\,\,}lr@{\,\,}lr@{\,\,}lr@{\,\,}l}
\hline\hline
\multicolumn{1}{c}{$N$} &
  \multicolumn{2}{c}{$\< R_e^2 \>_N$}  &
  \multicolumn{2}{c}{$\< R_g^2 \>_N$}  &
  \multicolumn{2}{c}{$\< R_m^2 \>_N$}  &
  \multicolumn{2}{c}{$F_N \< R_e^2 \>_N$}    \\
\hline
40 &   200.106 & $\pm$ 0.011 &
 & &
 & &
 & \\
60 &  364.977 & $\pm$ 0.012 &
     51.2026 & $\pm$ 0.0031 &
     160.10 & $\pm$ 0.04 &
     0.71 & $\pm$ 0.08 \\
80 &  559.656 & $\pm$ 0.033 &
     78.4660 & $\pm$ 0.0051 &
     245.51 & $\pm$ 0.06 &
     0.91 & $\pm$ 0.13  \\
100 &   780.245 & $\pm$ 0.028 &
     109.351 & $\pm$ 0.007 &
     342.34 & $\pm$ 0.09 &
     1.06 & $\pm$ 0.18 \\
120 &  1023.812 & $\pm$ 0.064 &
     143.488 & $\pm$ 0.010 &
      449.39 & $\pm$ 0.12 &
     1.02 & $\pm$ 0.24 \\
140 &  1288.631 & $\pm$ 0.064 &
     180.599 & $\pm$ 0.013 &
      565.93 & $\pm$ 0.16 &
     0.67 & $\pm$ 0.32 \\
150 &  1428.367 & $\pm$ 0.035 &
     200.185 & $\pm$ 0.015 &
      627.51 & $\pm$ 0.18 &
     0.32 & $\pm$ 0.36 \\
\hline\hline
\end{tabular}
\caption{Our Monte Carlo data for radii $\<R^2\>_N$
         as a function of walk length $N$.
         Errors (one standard deviation) are shown in parentheses.
}
\label{tableMC1}
\end{table}

\begin{table}
\centering
\begin{tabular}{rllll}\hline\hline
\multicolumn{1}{c}{$N$}  &  \multicolumn{1}{c}{$A_N$}  &
    \multicolumn{1}{c}{$B_N$}  &  \multicolumn{1}{c}{$C_N$}  &
    \multicolumn{1}{c}{$F_N$} \\
\hline
60 &  0.140290  (10) &
       0.438644  (100) &
       0.319826  (76) &
       0.001957  (200) \\
80 &  0.140204  (12) &
       0.438689  (120) &
       0.319598  (85) &
       0.001635  (230) \\
100 &  0.140149  (11) &
       0.438756  (120) &
       0.319425  (88) &
       0.001354  (240) \\
120 &  0.140150  (13) &
       0.438938  (120) &
       0.319294  (88) &
       0.000992  (240) \\
140 &  0.140148  (12) &
       0.439171  (130) &
       0.319118  (93) &
       0.000518  (250) \\
150 &  0.140149  (11) &
       0.439320  (130) &
       0.319014  (94) &
       0.000225  (250) \\
180 &  0.140161  (12) &
       0.439347  (71) &
       0.319022  (55) &
       0.000204  (140) \\
200 &  0.140171  (13) &
       0.439344  (63) &
       0.319047  (49) &
       0.000236  (120) \\
250 &  0.140171  (9) &
       0.439365  (64) &
       0.319030  (48) &
       0.000193  (130) \\
300 &  0.140192  (8) &
       0.439402  (26) &
       0.319052  (18) &
       0.000177  (45) \\
400 &  0.140208  (8) &
       0.439462  (26) &
       0.319045  (18) &
       0.000101  (46) \\
500 &  0.140216  (10) &
       0.439490  (33) &
       0.319043  (29) &
       0.000066  (66) \\
700 &  0.140242  (10) &
      & & \\
1000 &  0.140259  (7) &
       0.439557  (27) &
       0.319091  (19) &
       0.000047  (48) \\
1400 &  0.140258  (13) &
      & & \\
2000 &  0.140293  (14) &
       0.439630  (56) &
       0.319115  (45) &
      $-$0.000008  (110) \\
3000 &  0.140314  (19) &
      & & \\
4000 &  0.140274  (16) &
       0.439499  (71) &
       0.319167  (57) &
       0.000204  (140) \\
\hline\hline
\end{tabular}
\caption{Our Monte Carlo data for amplitude ratios
         as a function of walk length.
         Errors (one standard deviation) are shown in parentheses.
        }
\label{tableMC2}
\end{table}

\begin{table}[p]
\centering
\hspace*{-30pt}
\begin{tabular}{rlllrrr}
\hline\hline
\multicolumn{1}{c}{$N_{\rm min}$} &
  \multicolumn{1}{c}{$a_e$}  &
  \multicolumn{1}{c}{$b_e$}  &
  \multicolumn{1}{c}{$\Delta$}  &
  \multicolumn{1}{c}{$\chi^2$} &
  \multicolumn{1}{c}{DF}     &
  \multicolumn{1}{c}{CL} \\
\hline
   40 &  0.770998(23) &  0.001644(16) &  0.855(3) &
                                40.9 &  16 &   0.1\% \\
   60 &  0.771054(25) &  0.001589(19) &  0.869(4) &
                                14.9 &  15 &  46.2\% \\
   80 &  0.771081(30) &  0.001559(26) &  0.878(7) &
                                12.2 &  14 &  59.0\% \\
  100 &  0.771106(32) &  0.001532(29) &  0.887(8) &
                                 8.4 &  13 &  82.0\% \\
  120 &  0.771122(37) &  0.001512(36) &  0.894(12) &
                                 7.6 &  12 &  81.8\% \\
  140 &  0.771140(38) &  0.001491(38) &  0.902(13) &
                                 4.8 &  11 &  93.9\% \\
  150 &  0.771140(39) &  0.001492(39) &  0.902(13) &
                                 4.8 &  10 &  90.2\% \\
\hline\hline
\end{tabular}
\caption{Fit $\<R^2_e\> = N^{2 \nu} [a_e + b_e (N/750)^\Delta ]$. DF is the
number of degrees of freedom and CL is the confidence level of the fit.
}
\label{tableR2evar}
\end{table}

\begin{table}[p]
\centering
\hspace*{-30pt}
\begin{tabular}{rlllrrr}
\hline\hline
\multicolumn{1}{c}{$N_{\rm min}$} &
  \multicolumn{1}{c}{$a_g$}  &
  \multicolumn{1}{c}{$b_g$}  &
  \multicolumn{1}{c}{$\Delta$}  &
  \multicolumn{1}{c}{$\chi^2$}     &
  \multicolumn{1}{c}{DF}     &
  \multicolumn{1}{c}{CL} \\
\hline
   60 &  0.108213(4) &  0.000142(3) &  1.037(8) &
                                19.9 &  15 &  17.5\% \\
   80 &  0.108209(5) &  0.000146(4) &  1.023(11) &
                                16.8 &  14 &  26.7\% \\
  100 &  0.108205(5) &  0.000151(5) &  1.007(15) &
                                14.2 &  13 &  35.9\% \\
  120 &  0.108205(6) &  0.000150(5) &  1.008(18) &
                                14.2 &  12 &  28.8\% \\
  140 &  0.108207(6) &  0.000148(6) &  1.018(22) &
                                13.5 &  11 &  25.9\% \\
  150 &  0.108207(7) &  0.000148(7) &  1.019(26) &
                                13.5 &  10 &  19.5\% \\
  180 &  0.108210(8) &  0.000145(8) &  1.033(34) &
                                13.1 &   9 &  15.7\% \\
  200 &  0.108208(9) &  0.000147(9) &  1.023(41) &
                                12.9 &   8 &  11.4\% \\
  250 &  0.108205(10) &  0.000151(11) &  1.000(53) &
                                12.5 &   7 &   8.5\% \\
  300 &  0.108208(11) &  0.000147(13) &  1.025(66) &
                                12.1 &   6 &   6.0\% \\
  400 &  0.108201(15) &  0.000155(17) &  0.966(100) &
                                11.5 &   5 &   4.2\% \\
  500 &  0.108198(19) &  0.000158(21) &  0.937(146) &
                                11.5 &   4 &   2.2\% \\
  700 &  0.108213(20) &  0.000146(19) &  1.161(286) &
                                10.6 &   3 &   1.4\% \\
 1000 &  0.107954(1213) &  0.000384(1200) &  0.206(779) &
                                 8.8 &   2 &   1.3\% \\
\hline\hline
\end{tabular}
\caption{Fit $\<R^2_g\> = N^{2 \nu} [a_g + b_g (N/750)^\Delta ]$. DF is the
number of degrees of freedom and CL is the confidence level of the fit.
}
\label{tableR2gvar}
\end{table}

\begin{table}[p]
\centering
\hspace*{-30pt}
\begin{tabular}{rlllrrr}
\hline\hline
\multicolumn{1}{c}{$N_{\rm min}$} &
  \multicolumn{1}{c}{$a_m$}  &
  \multicolumn{1}{c}{$b_m$}  &
  \multicolumn{1}{c}{$\Delta$}  &
  \multicolumn{1}{c}{$\chi^2$}  &
  \multicolumn{1}{c}{DF}     &
  \multicolumn{1}{c}{CL} \\
\hline
   60 &  0.338957(32) &  0.000635(27) &  0.850(17) &
                                15.4 &  12 &  21.9\% \\
   80 &  0.338932(38) &  0.000661(33) &  0.828(23) &
                                13.3 &  11 &  27.1\% \\
  100 &  0.338940(41) &  0.000652(38) &  0.836(29) &
                                13.1 &  10 &  21.7\% \\
  120 &  0.338971(43) &  0.000618(41) &  0.871(36) &
                                10.2 &   9 &  33.6\% \\
  140 &  0.339008(44) &  0.000577(42) &  0.917(42) &
                                 5.7 &   8 &  68.1\% \\
  150 &  0.339018(45) &  0.000566(44) &  0.930(47) &
                                 5.3 &   7 &  62.7\% \\
  180 &  0.339013(48) &  0.000571(48) &  0.923(52) &
                                 5.2 &   6 &  52.1\% \\
  200 &  0.338990(56) &  0.000598(58) &  0.890(62) &
                                 4.2 &   5 &  51.7\% \\
  250 &  0.338962(71) &  0.000630(75) &  0.852(79) &
                                 3.7 &   4 &  45.3\% \\
  300 &  0.338955(75) &  0.000638(81) &  0.842(84) &
                                 3.6 &   3 &  31.3\% \\
  400 &  0.338915(115) &  0.000683(126) &  0.784(136) &
                                 3.3 &   2 &  19.2\% \\
  500 &  0.338896(146) &  0.000702(155) &  0.752(186) &
                                 3.2 &   1 &   7.2\% \\
\hline\hline
\end{tabular}
\caption{Fit $\<R^2_m\> = N^{2 \nu} [a_m + b_m (N/750)^\Delta ]$. DF is the
number of degrees of freedom and CL is the confidence level of the fit.
}
\label{tableR2mvar}
\end{table}

\begin{table}[p]
\centering
\hspace*{-30pt}
\begin{tabular}{rlllrrr}
\hline\hline
\multicolumn{1}{c}{$N_{\rm min}$} &
  \multicolumn{1}{c}{$a_e$}  &
  \multicolumn{1}{c}{$b_e$}  &
  \multicolumn{1}{c}{$c_e$}  &
  \multicolumn{1}{c}{$\chi^2$}  &
  \multicolumn{1}{c}{DF}     &
  \multicolumn{1}{c}{CL} \\
\hline
   40 &  0.771261(16) &     1.093(7) &  $-$1.93(4) &
                                15.4 &  16 &  49.4\% \\
   60 &  0.771247(18) &     1.104(9) &  $-$2.03(7) &
                                12.1 &  15 &  67.1\% \\
   80 &  0.771224(20) &     1.131(15) &  $-$2.30(14) &
                                 6.7 &  14 &  94.6\% \\
  100 &  0.771224(22) &     1.131(18) &  $-$2.30(18) &
                                 6.7 &  13 &  91.7\% \\
  120 &  0.771211(25) &     1.150(26) &  $-$2.52(29) &
                                 5.7 &  12 &  93.0\% \\
  140 &  0.771218(27) &     1.139(29) &  $-$2.39(32) &
                                 5.0 &  11 &  93.1\% \\
  150 &  0.771216(27) &     1.143(30) &  $-$2.44(34) &
                                 4.8 &  10 &  90.1\% \\
\hline\hline
\end{tabular}
\caption{Fit $\<R^2_e\> = N^{2 \nu} (a_e + b_e N^{-1}  + c_e N^{-3/2})$. DF is
the
number of degrees of freedom and CL is the confidence level of the fit.
}
\label{tableR2efixed}
\end{table}

\begin{table}[p]
\centering
\hspace*{-30pt}
\begin{tabular}{rlllrrr}
\hline\hline
\multicolumn{1}{c}{$N_{\rm min}$} &
  \multicolumn{1}{c}{$a_g$}  &
  \multicolumn{1}{c}{$b_g$}  &
  \multicolumn{1}{c}{$c_g$}  &
  \multicolumn{1}{c}{$\chi^2$}  &
  \multicolumn{1}{c}{DF}     &
  \multicolumn{1}{c}{CL} \\
\hline
   60 &  0.108209(3) &     0.107(2) &   0.08(2) &
                                17.8 &  15 &  27.5\% \\
   80 &  0.108207(4) &     0.109(3) &   0.06(3) &
                                16.3 &  14 &  29.3\% \\
  100 &  0.108204(4) &     0.113(4) &   0.02(4) &
                                14.3 &  13 &  35.6\% \\
  120 &  0.108204(5) &     0.113(5) &   0.02(5) &
                                14.3 &  12 &  28.5\% \\
  140 &  0.108206(5) &     0.110(6) &   0.05(6) &
                                13.5 &  11 &  25.9\% \\
  150 &  0.108206(5) &     0.110(6) &   0.06(8) &
                                13.5 &  10 &  19.6\% \\
  180 &  0.108208(6) &     0.106(8) &   0.11(11) &
                                13.0 &   9 &  16.4\% \\
  200 &  0.108207(7) &     0.108(10) &   0.08(14) &
                                12.9 &   8 &  11.7\% \\
  250 &  0.108205(8) &     0.113(13) &   0.00(19) &
                                12.5 &   7 &   8.5\% \\
  300 &  0.108208(9) &     0.106(17) &   0.12(25) &
                                12.0 &   6 &   6.1\% \\
  400 &  0.108203(11) &     0.120(26) &  $-$0.14(45) &
                                11.5 &   5 &   4.2\% \\
  500 &  0.108201(13) &     0.128(37) &  $-$0.30(72) &
                                11.5 &   4 &   2.2\% \\
  700 &  0.108212(17) &     0.076(64) &   0.95(1.46) &
                                10.5 &   3 &   1.5\% \\
 1000 &  0.108185(30) &     0.227(156) &  $-$3.03(4.03) &
                                 9.4 &   2 &   0.9\% \\
\hline\hline
\end{tabular}
\caption{Fit $\<R^2_g\> = N^{2 \nu} (a_g + b_g N^{-1}  + c_g N^{-3/2})$. DF is
the
number of degrees of freedom and CL is the confidence level of the fit.
}
\label{tableR2gfixed}
\end{table}

\begin{table}[p]
\centering
\hspace*{-30pt}
\begin{tabular}{rlllrrr}
\hline\hline
\multicolumn{1}{c}{$N_{\rm min}$} &
  \multicolumn{1}{c}{$a_m$}  &
  \multicolumn{1}{c}{$b_m$}  &
  \multicolumn{1}{c}{$c_m$}  &
  \multicolumn{1}{c}{$\chi^2$}  &
  \multicolumn{1}{c}{DF}     &
  \multicolumn{1}{c}{CL} \\
\hline
   60 &  0.339040(21) &     0.447(14) &  $-$0.99(12) &
                                18.6 &  12 &   9.9\% \\
   80 &  0.339013(23) &     0.479(18) &  $-$1.37(18) &
                                11.3 &  11 &  41.7\% \\
  100 &  0.339008(25) &     0.487(23) &  $-$1.47(26) &
                                11.0 &  10 &  35.7\% \\
  120 &  0.339019(27) &     0.469(28) &  $-$1.23(34) &
                                 9.8 &   9 &  36.4\% \\
  140 &  0.339038(29) &     0.436(33) &  $-$0.76(42) &
                                 6.3 &   8 &  61.7\% \\
  150 &  0.339044(30) &     0.426(37) &  $-$0.61(48) &
                                 5.9 &   7 &  55.6\% \\
  180 &  0.339041(32) &     0.431(41) &  $-$0.69(55) &
                                 5.8 &   6 &  44.8\% \\
  200 &  0.339028(34) &     0.457(49) &  $-$1.09(69) &
                                 4.9 &   5 &  43.1\% \\
  250 &  0.339010(41) &     0.496(68) &  $-$1.69(1.01) &
                                 4.2 &   4 &  38.0\% \\
  300 &  0.339005(43) &     0.508(75) &  $-$1.88(1.13) &
                                 4.1 &   3 &  25.5\% \\
  400 &  0.338976(62) &     0.583(140) &  $-$3.21(2.37) &
                                 3.6 &   2 &  16.1\% \\
  500 &  0.338964(74) &     0.623(194) &  $-$4.04(3.62) &
                                 3.6 &   1 &   5.9\% \\
\hline\hline
\end{tabular}
\caption{Fit $\<R^2_m\> = N^{2 \nu} (a_m + b_m N^{-1}  + c_m N^{-3/2})$. DF is
the
number of degrees of freedom and CL is the confidence level of the fit.
}
\label{tableR2mfixed}
\end{table}

\begin{table}[p]
\centering
\hspace*{-30pt}
\begin{tabular}{rlllrrr}
\hline\hline
\multicolumn{1}{c}{$N_{\rm min}$} &
  \multicolumn{1}{c}{$a_A$}  &
  \multicolumn{1}{c}{$b_A$}  &
  \multicolumn{1}{c}{$c_A$}  &
  \multicolumn{1}{c}{$\chi^2$}  &
  \multicolumn{1}{c}{DF}     &
  \multicolumn{1}{c}{CL} \\
\hline
   60 &  0.140305(6) &    $-$0.062(3) &   0.47(2) &
                                 8.4 &  15 &  90.7\% \\
   80 &  0.140306(6) &    $-$0.064(5) &   0.49(4) &
                                 8.1 &  14 &  88.2\% \\
  100 &  0.140303(7) &    $-$0.060(6) &   0.45(6) &
                                 7.1 &  13 &  89.6\% \\
  120 &  0.140306(8) &    $-$0.064(8) &   0.49(9) &
                                 6.6 &  12 &  88.2\% \\
  140 &  0.140306(8) &    $-$0.064(9) &   0.50(10) &
                                 6.6 &  11 &  83.0\% \\
  150 &  0.140307(9) &    $-$0.065(10) &   0.52(12) &
                                 6.5 &  10 &  77.0\% \\
  180 &  0.140309(10) &    $-$0.071(13) &   0.59(17) &
                                 6.2 &   9 &  72.5\% \\
  200 &  0.140310(11) &    $-$0.072(16) &   0.62(22) &
                                 6.1 &   8 &  63.4\% \\
  250 &  0.140307(12) &    $-$0.065(21) &   0.51(31) &
                                 5.9 &   7 &  55.7\% \\
  300 &  0.140312(14) &    $-$0.078(27) &   0.72(42) &
                                 5.3 &   6 &  50.3\% \\
  400 &  0.140310(18) &    $-$0.072(42) &   0.61(74) &
                                 5.3 &   5 &  38.2\% \\
  500 &  0.140304(21) &    $-$0.049(59) &   0.12(1.15) &
                                 5.0 &   4 &  28.9\% \\
  700 &  0.140306(27) &    $-$0.060(104) &   0.39(2.38) &
                                 5.0 &   3 &  17.4\% \\
 1000 &  0.140297(47) &    $-$0.007(244) &  $-$1.03(6.34) &
                                 4.9 &   2 &   8.6\% \\
\hline\hline
\end{tabular}
\caption{Fit $A_N = a_A + b_A N^{-1}  + c_A N^{-3/2}$. DF is the
number of degrees of freedom and CL is the confidence level of the fit.
}
\label{tableAfixed}
\end{table}

\begin{table}[p]
\centering
\hspace*{-30pt}
\begin{tabular}{rlllrrr}
\hline\hline
\multicolumn{1}{c}{$N_{\rm min}$} &
  \multicolumn{1}{c}{$a_B$}  &
  \multicolumn{1}{c}{$b_B$}  &
  \multicolumn{1}{c}{$c_B$}  &
  \multicolumn{1}{c}{$\chi^2$}  &
  \multicolumn{1}{c}{DF}     &
  \multicolumn{1}{c}{CL} \\
\hline
   60 &  0.439619(31) &    $-$0.062(21) &  $-$0.02(17) &
                                12.8 &  12 &  38.1\% \\
   80 &  0.439587(34) &    $-$0.024(27) &  $-$0.45(27) &
                                 8.3 &  11 &  68.3\% \\
  100 &  0.439578(38) &    $-$0.012(34) &  $-$0.61(37) &
                                 8.0 &  10 &  63.2\% \\
  120 &  0.439596(41) &    $-$0.040(42) &  $-$0.24(49) &
                                 6.6 &   9 &  67.5\% \\
  140 &  0.439620(43) &    $-$0.082(49) &   0.34(61) &
                                 4.0 &   8 &  85.6\% \\
  150 &  0.439629(45) &    $-$0.098(55) &   0.59(70) &
                                 3.5 &   7 &  83.4\% \\
  180 &  0.439626(48) &    $-$0.091(61) &   0.48(80) &
                                 3.4 &   6 &  75.2\% \\
  200 &  0.439614(51) &    $-$0.068(73) &   0.12(1.01) &
                                 3.1 &   5 &  68.5\% \\
  250 &  0.439590(61) &    $-$0.014(102) &  $-$0.72(1.52) &
                                 2.5 &   4 &  63.6\% \\
  300 &  0.439583(65) &     0.004(118) &  $-$1.02(1.79) &
                                 2.4 &   3 &  48.5\% \\
  400 &  0.439558(89) &     0.073(207) &  $-$2.27(3.55) &
                                 2.3 &   2 &  31.9\% \\
  500 &  0.439528(106) &     0.174(284) &  $-$4.35(5.35) &
                                 2.0 &   1 &  15.6\% \\
\hline\hline
\end{tabular}
\caption{Fit $B_N = a_B + b_B N^{-1}  + c_B N^{-3/2}$. DF is the
number of degrees of freedom and CL is the confidence level of the fit.
}
\label{tableBfixed}
\end{table}

\begin{table}[p]
\centering
\hspace*{-30pt}
\begin{tabular}{rlllrrr}
\hline\hline
\multicolumn{1}{c}{$N_{\rm min}$} &
  \multicolumn{1}{c}{$a_F$}  &
  \multicolumn{1}{c}{$b_F$}  &
  \multicolumn{1}{c}{$c_F$}  &
  \multicolumn{1}{c}{$\chi^2$}  &
  \multicolumn{1}{c}{DF}     &
  \multicolumn{1}{c}{CL} \\
\hline
   60 &  0.000043(59) &    $-$0.039(39) &   1.30(33) &
                                11.5 &  12 &  49.0\% \\
   80 &  0.000105(65) &    $-$0.112(52) &   2.14(51) &
                                 6.7 &  11 &  81.9\% \\
  100 &  0.000117(71) &    $-$0.129(65) &   2.36(71) &
                                 6.5 &  10 &  76.8\% \\
  120 &  0.000087(77) &    $-$0.082(79) &   1.72(94) &
                                 5.4 &   9 &  79.4\% \\
  140 &  0.000041(82) &    $-$0.004(93) &   0.64(1.17) &
                                 3.0 &   8 &  93.5\% \\
  150 &  0.000026(86) &     0.023(104) &   0.24(1.34) &
                                 2.6 &   7 &  91.7\% \\
  180 &  0.000037(90) &     0.002(116) &   0.54(1.54) &
                                 2.5 &   6 &  87.2\% \\
  200 &  0.000064(98) &    $-$0.052(139) &   1.38(1.94) &
                                 2.0 &   5 &  85.3\% \\
  250 &  0.000110(116) &    $-$0.153(195) &   2.95(2.89) &
                                 1.4 &   4 &  83.9\% \\
  300 &  0.000133(124) &    $-$0.205(220) &   3.80(3.32) &
                                 1.2 &   3 &  76.2\% \\
  400 &  0.000167(174) &    $-$0.298(398) &   5.45(6.75) &
                                 1.1 &   2 &  58.1\% \\
  500 &  0.000189(206) &    $-$0.372(545) &   6.98(10.23) &
                                 1.0 &   1 &  30.7\% \\
\hline\hline
\end{tabular}
\caption{Fit $F_N = a_F + b_F N^{-1}  + c_F N^{-3/2}$. DF is the
number of degrees of freedom and CL is the confidence level of the fit.
}
\label{tableFfixed}
\end{table}

\begin{table}[p]
\centering
\hspace*{-30pt}
\begin{tabular}{rlrrr}
\hline\hline
\multicolumn{1}{c}{$N_{\rm min}$} &
  \multicolumn{1}{c}{$f$}  &
  \multicolumn{1}{c}{$\chi^2$}  &
  \multicolumn{1}{c}{DF}     &
  \multicolumn{1}{c}{CL} \\
\hline
   60 &       0.75(5) &  12.7 &  14 &  55.4\% \\
   80 &       0.79(7) &  12.2 &  13 &  51.3\% \\
  100 &       0.73(8) &  10.7 &  12 &  55.4\% \\
  120 &       0.65(9) &   6.7 &  11 &  82.0\% \\
  140 &       0.58(10) &   4.1 &  10 &  94.5\% \\
  150 &       0.57(11) &   4.0 &   9 &  91.2\% \\
  180 &       0.60(11) &   3.5 &   8 &  90.3\% \\
  200 &       0.64(12) &   2.7 &   7 &  91.5\% \\
  250 &       0.68(14) &   2.4 &   6 &  88.1\% \\
  300 &       0.69(15) &   2.3 &   5 &  80.2\% \\
  400 &       0.64(25) &   2.3 &   4 &  68.6\% \\
  500 &       0.69(51) &   2.3 &   3 &  52.0\% \\
 1000 &       1.18(1.17) &   2.0 &   2 &  36.1\% \\
\hline\hline
\end{tabular}
\caption{Fit $F_N \<R^2_e\>_N = f$. DF is the
number of degrees of freedom and CL is the confidence level of the fit.
}
\label{tableFRE}
\end{table}

\end{document}